\documentclass[%
 aip,
 amsmath,amssymb,
 reprint,%
]{revtex4-1}

\usepackage{graphicx}
\usepackage{dcolumn}
\usepackage{bm}
\usepackage[usenames,dvipsnames]{color}
\usepackage[utf8]{inputenc}
\usepackage[T1]{fontenc}
\usepackage{mathptmx}
\usepackage{etoolbox}

\makeatletter
\newcommand{\ostar}{\mathbin{\mathpalette\make@circled\star}}

\def\@email#1#2{%
 \endgroup
 \patchcmd{\titleblock@produce}
  {\frontmatter@RRAPformat}
  {\frontmatter@RRAPformat{\produce@RRAP{*#1\href{mailto:#2}{#2}}}\frontmatter@RRAPformat}
  {}{}
}%
\makeatother
\begin{document}

\preprint{AIP/123-QED}

\title[Towards machine learning for microscopic mechanisms: a formula search for crystal structure stability based on atomic properties]{Towards machine learning for microscopic mechanisms:\\ a formula search for crystal structure stability based on atomic properties}
\author{Udaykumar Gajera}
\affiliation{Consiglio Nazionale delle Ricerche, CNR-SPIN c/o Università ``G. D’Annunzio", 66100 Chieti, Italy}
\affiliation{Chemistry Department, University of Turin, via Pietro Giuria, 7, 10125, Torino, Italy}

\author{Loriano Storchi}%
\affiliation{ 
Dipartimento di Farmacia, Universitá degli Studi G. D’Annunzio, 66100 Chieti, Italy
}%

\author{Danila Amoroso}
\affiliation{Consiglio Nazionale delle Ricerche, CNR-SPIN c/o Università ``G. D’Annunzio", 66100 Chieti, Italy}
 \affiliation{ NanoMat/Q-mat/CESAM,Universite de Liege, B-4000 Liege, Belgium}

\author{Francesco Delodovici}
\affiliation{Consiglio Nazionale delle Ricerche, CNR-SPIN c/o Università ``G. D’Annunzio", 66100 Chieti, Italy}

\author{Silvia Picozzi}
\affiliation{Consiglio Nazionale delle Ricerche, CNR-SPIN c/o Università ``G. D’Annunzio", 66100 Chieti, Italy}

\date{\today}

\begin{abstract}
Machine Learning (ML) techniques are revolutionizing the way to perform efficient materials modeling. Nevertheless, not all the ML approaches allow for the understanding of microscopic mechanisms at play in different phenomena. To address the latter aspect, we propose a combinatorial machine-learning approach to obtain physical formulas based on simple and easily-accessible ingredients, such as atomic properties. The latter are used to build materials features that are finally employed, through Linear Regression, to predict the energetic stability of semiconducting binary compounds with respect to zincblende and rocksalt crystal structures. The adopted models are trained using dataset built from first-principles calculations. Our results show that already one-dimensional (1D) formulas
well describe the energetics; a simple grid-search optimization of the automatically-obtained 1D-formulas enhances the prediction performances at a very small
computational cost. In addition, our approach
allows to highlight the role of the different atomic properties involved
in the formulas. The computed formulas clearly indicate that ``spatial" atomic properties ({\em i.e.} radii indicating maximum probability densities for $s,p,d$ electronic shells) drive the stabilization of one crystal structure with respect to the other, suggesting the major relevance of the radius associated to the $p$-shell of the cation species. 
\end{abstract}

\maketitle

\section{Introduction}


 Modeling material properties with high accuracy and low computational cost is one of the grand-challenges in materials science and engineering. 
The development of ab-initio methods have provided accurate tools for material properties prediction and their further optimization; nevertheless, one disadvantage of approaches  relying only on first-principles simulations is the high cost required in terms of computational resources and simulation time. In recent years, the continuous growth of available computational power \cite{moore_cramming_1965} has stimulated scientists to move in the direction of high-throughput simulations\cite{ramprasad_machine_2017,fukuda_structure_2021,Schwalbe-Koda_2021,Homer_2019,curtarolo_high-throughput_2013, green_fulfilling_2017, walsh_quest_2015, shen_high-throughput_2021, griesemer_high-throughput_2021}. Along this line, open access databases, such as OQMD\cite{saal_materials_2013}\cite{kirklin_open_2015}, NOMAD\cite{draxl_nomad_2018, draxl_nomad_2019}, Aflowlib\cite{curtarolo_aflowliborg_2012}, C2DB\cite{gjerding_recent_2021, haastrup_computational_2018}, QPOD\cite{bertoldo_quantum_2021}, Materials Project\cite{jain_commentary_2013}, Materials Cloud\cite{talirz_materials_2020} and related AiiDa\cite{pizzi_aiida_2016,huber_common_2021}, provide researchers with a huge collection of basic first-principles results. A large amount of ab-initio data is thus available, which can be used for deeper analyses and studies, provided one can count on proper tools to extract relevant 
information out of them. 
%
  Therefore, in the last years, materials scientists have developed different Machine Learning (ML) methods to rationalize the data analysis
  \cite{park_data-driven_2021,kim_organized_2016,tsymbalov_machine_2021,bartel_new_2019,kusne_--fly_2014,koinuma_combinatorial_2004, manti_predicting_2022, kim_machine-learning-accelerated_2018, pal_scale-invariant_2021, kuban_density--states_2022}. Each method has its own specific advantages and limitations. Methods like Neural Network (NN)\cite{gurney_introduction_0000} or Random Forest ~\cite{breiman_random_2001} are very efficient\cite{xie_crystal_2018} but not always transparent, blurring the comprehension of the role played by the input variables in the final results;  
 %
 ML methods, based for instance on linear regression (LR) \cite{chatterjee_handbook_2013,leskovec_mining_nodate}, appear to be more suitable to obtain predictive and comprehensible models \cite{miller2019explanation,kim2016examples}.
  Nevertheless, finding a linear dependence between input and output properties is not always an easy task.
  
  In this work, we thus propose a ML-based approach to build sets of features (or descriptors) starting from a given set of basic variables (e.g., atomic properties), which are subsequently used to construct LR models (or formulas). 
    To test our method, we target a prototypical case in material science: the classification of the most stable crystal structure between rock-salt (RS) and zinc-blende (ZB) for semiconductor AB binary compounds \cite{ghiringhelli_big_2015}. In our approach, we adopt both simple 
one-dimensional and multi-dimension LR. 
%
To identify useful features, we generate combinations of basic atomic properties ({\em i.e.} the independent variables in our approach) of the material constituents through a combinatorial approach\cite{meredig_combinatorial_2014}. 
We then carry out an analysis of the emerging best-performing formulas, identifying the role of specific  atomic features in determining the final stabilization of the crystal structure. 
Finally, we test the predictive capability of the obtained formulas by applying them to ``new" compounds ({\em i.e.} outside the dataset used for training the model), finding an overall satisfactory agreement with first-principles results. We remark 
that our approach is similar to what originally  proposed by Ghiringhelli {\em et al.}~\cite{ghiringhelli_big_2015}, though with some differences and further extensions,  which will be carefully discussed in what follows.




\color{black}
\section{Methodology}\label{sec:method}

The approach we present here can be regarded as a combinatorial machine-learning: a set 
of basic atomic properties (APs, listed in Table S.2 in Supplementary Information) are randomly combined (though under certain initial constraints detailed below), to build a set of material features (MFs). The generated features are then
used to train a LR model, where the energy difference between rocksalt and zincblende structures is the dependent variable (i.e., the label). Then, we select the best performing model according to standard performances metrics, such as the Root Mean Squared Error (RMSE). The final result of this procedure is a ``formula'', which is a concise and clear representation of the relationship between the used atomic properties and the energy difference between RS and ZB phases. In the following, we describe in detail the different steps of our approach.


\subsection{Dataset preparation and materials} \label{DP}

As mentioned, we aim at predicting the total energy difference ($\Delta E = E^{RS}-E^{ZB}$) between RS and ZB phases of cubic crystal structures for 82 semiconductor binary AB compounds
(the dataset is reported in table S.2 in SI). We employed total energies reported in Ref.~\cite{ghiringhelli_big_2015}, which were calculated through density functional theory (DFT)\cite{DFT}\cite{DFT3} within the local density approximation (LDA~\cite{DFT2}).

The construction of the material features (MFs), is based on primary atomic properties  
of the constituents, also taken from Ref. \cite{ghiringhelli_big_2015}. To facilitate the physical interpretation of each MF, the APs are subdivided into two different kinds:
($i$) ``energy" properties, including highest occupied Kohn-Sham level (HOMO), lowest unoccupied Kohn-Sham level (LUMO), Ionization Potential (IP), Electron affinity (EA); ($ii$) ``spatial" properties, including $r_s$, $r_p$, and $r_d$, {\em i.e.} the radii where the radial probability density of the valence $s$, $p$, and $d$ orbitals, respectively, reaches its maximum. 

\begin{figure*}
    \centering
     
    \includegraphics[scale=0.38]{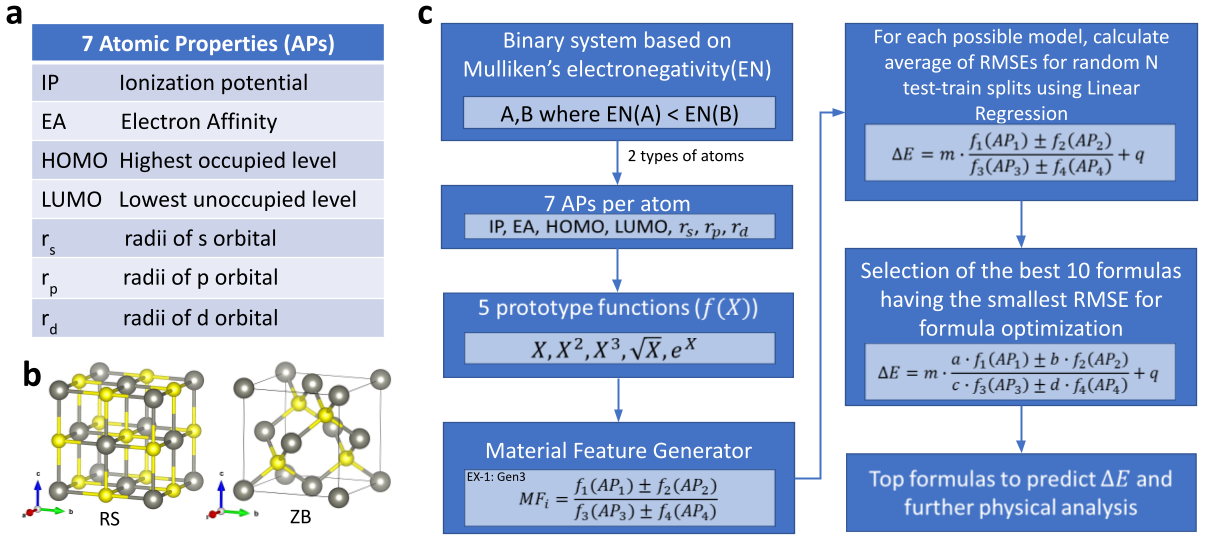}
    \caption{a) Basic atomic properties (APs) used to construct the material features. 
    b) Crystal structures of RS and ZB (plot made using the VESTA tool) \cite{Momma:vesta}. Grey (yellow) spheres represent A (B) atoms. c) Workflow for formulas construction, machine learning methodology, validation, and MF selection procedures. In the AB compounds, A is the atom with the lowest electronegativity. } 
    \label{fig:workflow}
\end{figure*}

\subsubsection{Formulas construction} \label{LR}
We rely on the LR\cite{chatterjee_handbook_2013}\cite{leskovec_mining_nodate}  approach to obtain a direct interpretation of the dependent and independent variables.
The construction of a useful LR model  can become troublesome, requiring a linear dependence between features. 
In Ref.\cite{ghiringhelli_big_2015}, the authors implemented an automated feature selection method employing the LASSO regression analysis method~\cite{shalev-shwartz_understanding_2014,ghiringhelli_big_2015}. In our work, we use a combinatorial approach to generate the dependent variable (material features) to be used within the linear equations, and thus to finally obtain the formulas. 

In Fig.~\ref{fig:workflow}, we illustrate the workflow of the formula generation and selection using LR. 
The process starts with the selection of the APs to be combined.
Afterwards, we choose prototype functions that are simple analytical operations applied to the APs. In our case, we selected 5 prototype functions, $f(x)$, namely $x, x^2, x^3, \sqrt{x}, e^x$. where $x$ is an AP. 
Then, we obtain the final set of MFs by combining different prototype functions via the combinatorial approach (see for instance \cite{meredig_combinatorial_2014}), and applying the following additional set of rules:


\begin{itemize}
    \item $GEN1$:  combine two prototype functions in the numerator, forcing them to belong to the same kind of APs, that is both ``spatial"-like or both ``energy"-like;  one prototype function is at the denominator with the only constraint to be non-zero, such as
    \begin{equation}\label{eq-gen1}
        MF = \frac{f_1(AP_1)\pm f_2(AP_2)}{f_3(AP_3)}
    \end{equation}
    
    \item $GEN2$:  combine two prototype functions with same kind of APs at the numerator, and a single prototype function at the denominator with argument of a different kind with respect to the numerator ones. For instance, if $AP_1$ in $f_1(AP_1)$ and $AP_2$ in $f_2(AP_2)$ is an ``energy" term ({\em i.e.} $EA$ or $HOMO$), then $AP_3$ must be a ``spatial" term, ({\em i.e.} $r_p$)
    \begin{equation}\label{eq-gen2}
        MF = \frac{f_1(AP_1)\pm f_2(AP_2)}{f_3(AP_3)}
    \end{equation}
    
    \item $GEN3$:  combine two  prototype functions at both the numerator and denominator without any constraints 
    \begin{equation}\label{eq-gen3}
        MF = \frac{f_1(AP_1)\pm f_2(AP_2)}{f_3(AP_3)\pm f_4(AP_4)}
    \end{equation}
    
    \item $GEN4$: combine two prototype functions with the same physical dimensions at both the numerator and denominator
    \begin{equation}\label{eq-gen4}
        MF = \frac{f_1(AP_1)\star f_2(AP_2)}{f_3(AP_3)\star  f_4(AP_4)}
    \end{equation}
    where $\star=+-\times\div$
    
\end{itemize}
Each one of these set of rules corresponds to a different MFs generator.

From the implementation point of view, each generator is a Python \cite{van1995python} function that produces a set of strings. Therefore, we can easily exploit the Python 
capability to parse a source code and run Python expression (code) within a program \cite{Storchi} to compute all the MFs' values starting from 
the generated sets of strings. This allows for an easy implementation and plugin of other generators, leaving the 
workflow unchanged: a new generator can be introduced implementing a Python function returning a list of strings, each one being a valid MF.

Finally, in order to choose the optimal formula, we build a LR  model for each of the generated MF. To practically select the best model, \emph{i.e.} the ``best formula'', we randomly split the full dataset into : 90\% as training set to train/initialize the model; 10\% as a test set to check model's performance. We perform this random splitting $N$ times (with $N=150$) for each model, and we calculate the RMSE from the test set for each run. Afterwards, we again verify the top 10 resulting best formulas with a higher value of training set and test set splitting, with $N=1000$. We average it over all $N$ splitting, and we obtain $avg(RMSE)$, as reported in our Tables.


We mention that different metrics for evaluating regression model can lead to different formulas ranking.
In this work, we rank the obtained models based on the lowest $avg(RMSE)$ for direct comparison with a previous work~\cite{ghiringhelli_big_2015}.

\subsection{Formula optimization}

In order to further improve the performance of our models, we introduce an additional step, which we refer to as ``formula optimization". In detail, we focus on the top 10 formulas obtained using each generator and the subsequent LR, as described in the previous section. After that, we use a grid search to find the relative weights of each prototype function of the atomic properties (i.e., each $f_i(AP_i)$) within the formula. A first grid search ranging between -1 to 1 with the increasing step of 0.1 is used. We multiply each $f_i(AP_i)$ of the formula by the weight coefficient and we optimize the final RMSE value.
Once the procedure finds a set of optimal weight coefficients, two subsequent grid searches, with reduced incremental step values (0.01 and 0.001 respectively) and range of search are performed to obtain the final set of refined weight coefficients. Noteworthy, for each set of weight coefficients generated during the grid search, we also run the linear regression. Thus, we are performing a proper formula optimization, as at each step of the grid search we are updating both the weight coefficients as well as the slope and intercept coming from the LR.  

 To further clarify the procedure, we show here an exemplary equation:
 \begin{equation}
    \Delta E = m \cdot \frac{a\cdot f_1(AP_1) \star b\cdot f_2(AP_2) }{c\cdot f_3(AP_3) \star d\cdot f_4(AP_4)} + q 
    \label{eq-1}
\end{equation}

where $\Delta E$ is the targeted material feature (MF), $a,b,c,d$ denote the weight coefficients scanned during the grid search, $f_1(AP_1),f_2(AP_2),f_3(AP_3),f_4(AP_4)$ are the prototype functions build on the primary atomic properties $AP_i$, and $m$ and $q$ are the the slope or angular coefficient and intercept, respectively,  recursively determined upon LR.

In Table~\ref{tab:1d_2}, we report the optimized, best performing formula from the different generators; the top 10 formulas are reported in Table S.1 of the Supplementary Material.

To benchmark our grid search, we also used automated coefficient-optimizing methods: Nelder-Mead\cite{gao_implementing_2012}, Conjugate Gradient (CG)\cite{golub_matrix_2013}, Broyden–Fletcher–Goldfarb–Shanno (BFGS) \cite{10.1093/imamat/6.1.76} and Truncated Newton method (TNC)\cite{dembo_inexact_1982}.  
Although the resulting sets of coefficients are different in terms of single values with respect to those obtained via the grid search, the ratios between them is almost preserved as well as the associated RMSE. In particular, for the case of $GEN1$ and $GEN2$, the ratio between the numerator coefficients $a$ and $b$ is preserved; for $GEN3$ and $GEN4$ also the denominator coefficients ratio, between $c$ and $d$, is preserved. 
In Fig. S.3 of the Supplementary Material, we show the evolution of the RMSE and different ratios for different methods using 1D feature generated by $GEN3$. 



\subsection{Higher-dimensional features}
For the construction of higher dimensional 2D-formulas, we combined in all possible ways pairs of MFs extracted from the best 1000 ones and checked the $avg(RMSE)$ using multiple LR  for $N$ test-train set splits. We followed the same process to construct the 3D formulas, where three different 1D MFs are combined. The comparison between performances is discussed in the following Section.

\subsection{Test of predictive power of $\Delta E$ formula for novel AB compounds} 

After obtaining the optimised 1D formulas for $\Delta E$ in the case of AB compounds, we aimed at further verifying their validity and  predictive power, by considering additional AB systems ({\em i.e.} which were not originally included in the ML training set) and by comparing values obtained from ML-predicted $\Delta E$ formula with  corresponding \emph{ab-initio} calculated values. 
In closer detail, we focused on different alloys, obtained by changing respectively the concentration of  A-site atoms, such as $[A_xA'_{1-x}]B$, and of  B-site atoms, such as $A[B_xB'_{1-x}]$. Accordingly, one can test the efficiency of the formulas by checking the energy difference for intermediate concentrations as obtained from optimised 1D formulas and compare their trend with respect to first-principles results. To this end, ab-initio electronic-structure simulations were carried out within DFT 
and LDA functional.
Calculations were performed using the VASP\cite{VASP1,VASP2,VASP3} code, employing a $8\times 8\times 8$ k-mesh for the Brillouin zone sampling. We verified that the results obtained with the pseudopotential VASP for the parent binary compounds were consistent with those reported by Ghiringhelli {\em et al.},  calculated with the all-electron FHI-aims code \cite{blum_ab_2009}. For simulations at different concentrations, we adopted the so called ``Virtual Crystal Approximation'' (VCA), based on virtual atoms interpolating between the real constituent atoms~\cite{VCA1,VCA2}. However, as well
known from the literature, the VCA approach neglects some effects, such as local distortions around atoms and, as such, should not be expected to reproduce fine details of disordered alloys properties~\cite{VCA_dan}.
Accordingly, in some cases ({\em i.e.} for Mg$_x$Ca$_{1-x}$Se alloys), 
in order to  mimic disordered structures with an improved accuracy, 
we calculated total energies using supercell structures, rather than using the VCA method on primitive unit cells. Specifically, 
the considered supercell is the cubic unit cell 
composed by four AB formula units with planes of cations alternating along the \textbf{c} direction (see Figure S.4). 
The $k$-mesh was modified accordingly, to maintain the same density of points employed in the simulations of primitive cells.

\section{Results and Discussion}

In this section, we will analyse the final formulas as obtained from different generators. The results are shown in Tables \ref{tab:1d},\ref{tab:1d_2},\ref{tab:2d},\ref{tab:3D}; in the first row we report the results obtained by Ghiringhelli {\em et al.}\cite{ghiringhelli_big_2015} for comparison. 

    
    First, by comparing the $avg(RMSE)$ values, we note that all 1D formulas obtained from our different generators better perform with respect to the 1D ones reported in \cite{ghiringhelli_big_2015}, where the authors used the automated feature selection method LASSO~\cite{shalev-shwartz_understanding_2014}. 
    Noteworthy, some atomic primary features appearing in 1D formulas of Ref. \cite{ghiringhelli_big_2015} also appear in our obtained list of 1D formulas using $GEN1$ and $GEN2$; nevertheless, those are characterized by a higher $avg(RMSE)$ than other formulas we obtained via our combinatorial approaches. 
    Additionally, formulas from $GEN3$ show the lowest $avg(RMSE)$ among all the others.   
    We also note, from
    Table~\ref{tab:1d}, that $GEN1$ and $GEN3$ provide lower $avg(RMSE)$ compared to $GEN2$ and $GEN4$ respectively; however, $GEN2$ and $GEN4$ have a higher success rate in terms of classification prediction. This testifies the fact that the choice of the performances metric to rank the material features can be different according to the target problem to be studied; different models' performances metric are, in fact, not always correlated.


    

In order to gather hints on the relative contribution of the individual primary atomic properties to the stabilization of either the rocksalt or the zincblende structure, 
we extracted the best ten formulas with the lowest $avg(RMSE)$ from each generator (so called ``original" formulas) and then apply the formula optimization, as detailed in the previous section. 
This procedure attributes relative weights to each $f(AP)$, allowing to measure the importance of the individual atomic properties in driving the energy stabilization.
In principle, the $avg(RMSE)$ value depends on random test-train splits that we perform to our dataset. 
Therefore, to reduce the effect of randomization, as a target model performances metric, we rank our optimized formula based on the RMSE of the whole dataset,  rather than based on $avg(RMSE)$. 
By comparing Table~\ref{tab:1d} and Table~\ref{tab:1d_2}, it is evident that the optimization procedure can further change the formulas ranking, providing a different final ``best formula'' with respect to the non-optimized formulas. In particular, we notice an improvement in RMSE around 5-10\% after the formula optimization. 

\begin{figure}[h!]
    \centering
    
    \includegraphics[scale=0.11]{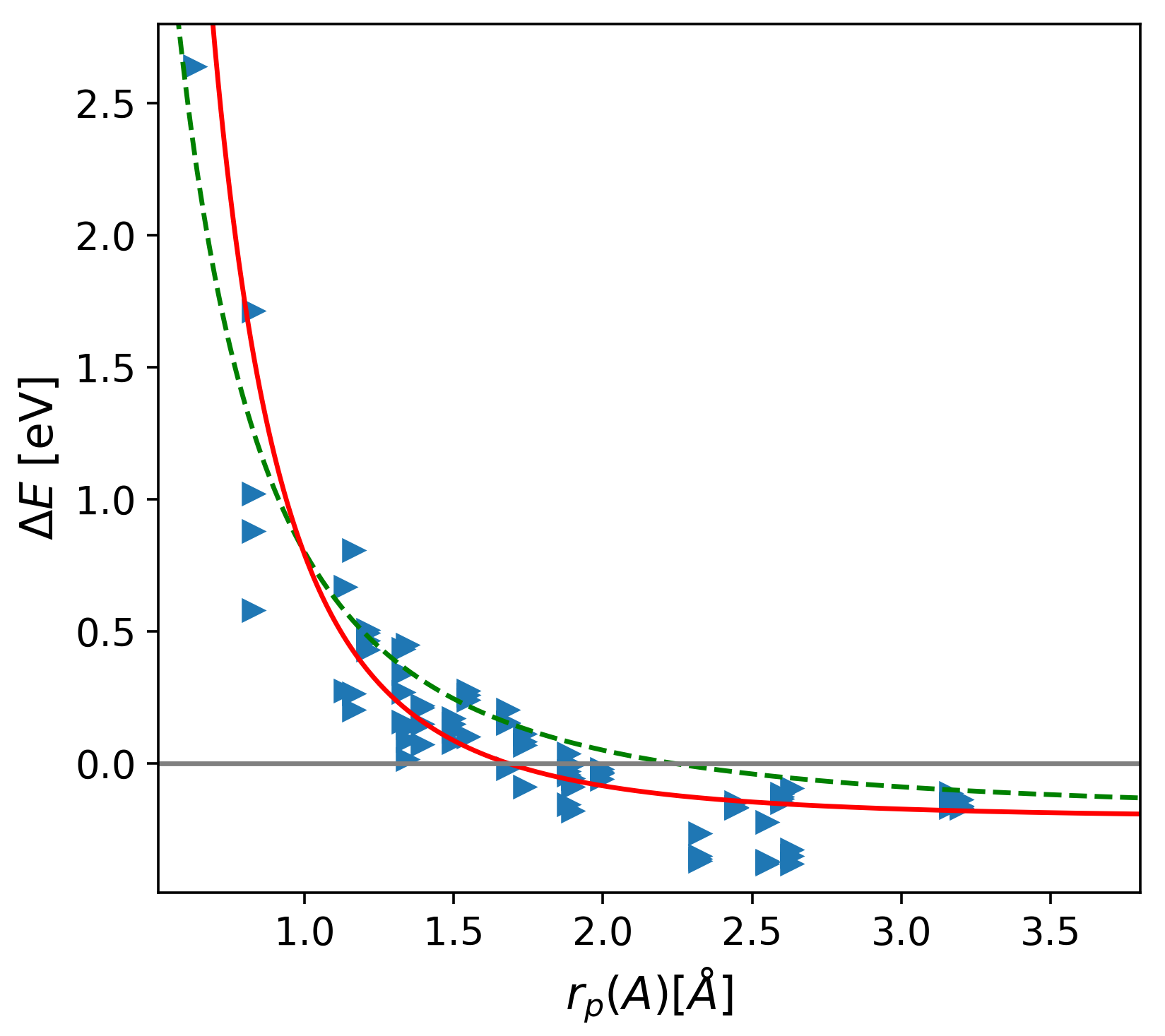}
    \caption{ Energy difference between rocksalt and zincblende, $\Delta E$ (in eV), as a function of $r_p(A)$ for different binary compounds (blue triangles). Data fit functions are also shown, using proportionality to $r_p(A)^{-2}$  and $r_p(A)^{-3}$ via green dashed line and red straight line, respectively. }
    \label{fig:rp_vs_de}
\end{figure}

Interestingly, our results reveal the size of the A-ion to play a leading role in the phase stabilization; in fact, the $r_p(A)$ radius appears in the best performing formulas more frequently than the other basic atomic properties. 
Therefore, we further analysed the dependence of $\Delta E$ on $r_p(A)$. In Fig.~\ref{fig:rp_vs_de}, we show $\Delta E$ as a function of $r_p(A)$, including fitting curves proportional to $r_p(A)^{-2}$  and $r_p(A)^{-3}$. What can be observed is a clear dependence of $\Delta E$ on $r_p(A)$: larger (smaller) $r_p(A)$ favors RS (ZB). 
Moreover, there is an overall good agreement with the fit, particularly using the $r_p(A)^{-3}$ function. The latter is, in fact, the most recurrent prototype function detected by the ML models.
Such a strong dependence for the energy is not observed with respect to the other atomic properties; other comparative plots of 
$\Delta E$ as a function of other $f(p)$ are  reported in Fig. S.2 of the Supplementary Material.

From the obtained results, we remark 
that formulas based on ``spatial'' atomic properties achieve higher ranking, thus better performance, with respect to those including atomic energy terms, both in the original models and in the optimized ones. Accordingly, this behaviour further confirms the primary role played by the atomic size (or, equivalently, steric effects), in determining the energetics of the AB compounds, \emph{i.e.} in selecting the preferred crystal structure \cite{ghiringhelli_big_2015}.

\begin{figure*}
    \centering
      
    \includegraphics[scale=0.25]{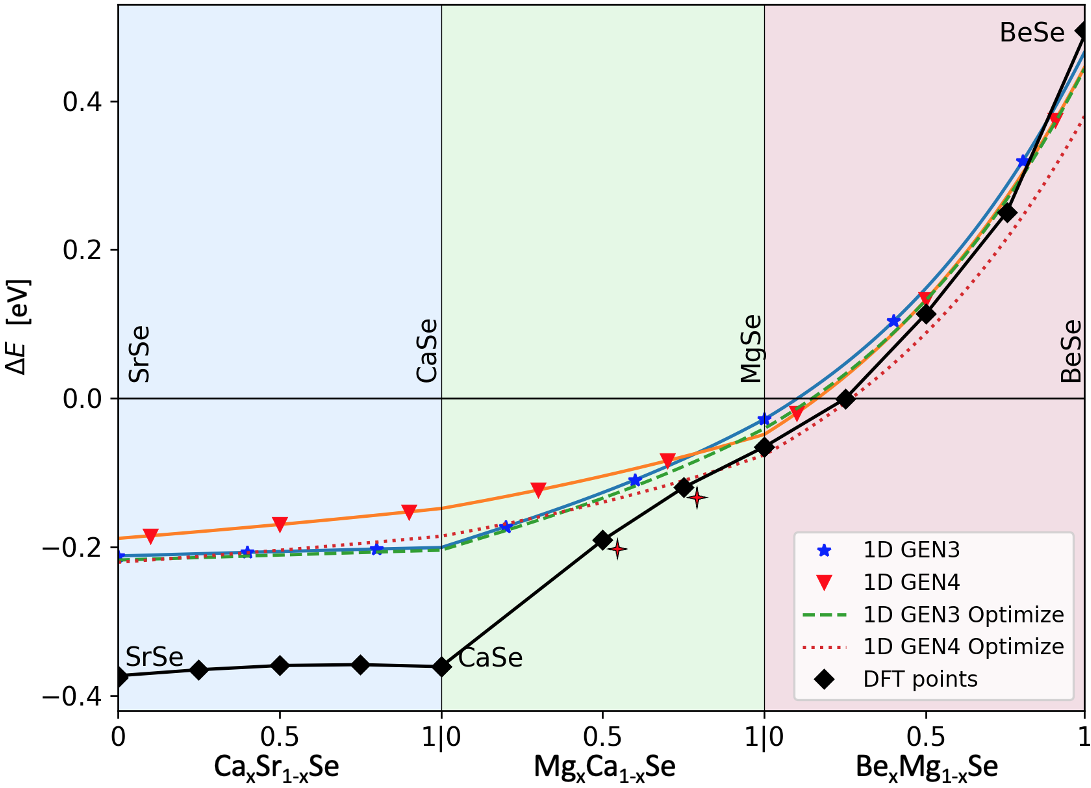}
    \caption{Total energy difference $\Delta E$ as a function of concentration($x$) for $[Ca_xSr_{1-x}]Se$, $[Mg_xCa_{1-x}]Se$ and $[Be_xMg_{1-x}]Se$ alloys, highlighted in blue, green, and pink regions respectively. Energy differences are predicted using original and optimized 1D descriptors constructed using $GEN3$ and $GEN4$ and verified using DFT (black line with diamond points) within VCA. For an improved accuracy, the two asterisk-highlighted intermediate points in the $[Mg_xCa_{1-x}]Se$ region are calculated using the supercell approach rather than  VCA.}
    \label{fig:a_site_change}
\end{figure*}

\begin{figure*}
    \centering
      
    \includegraphics[scale=0.43]{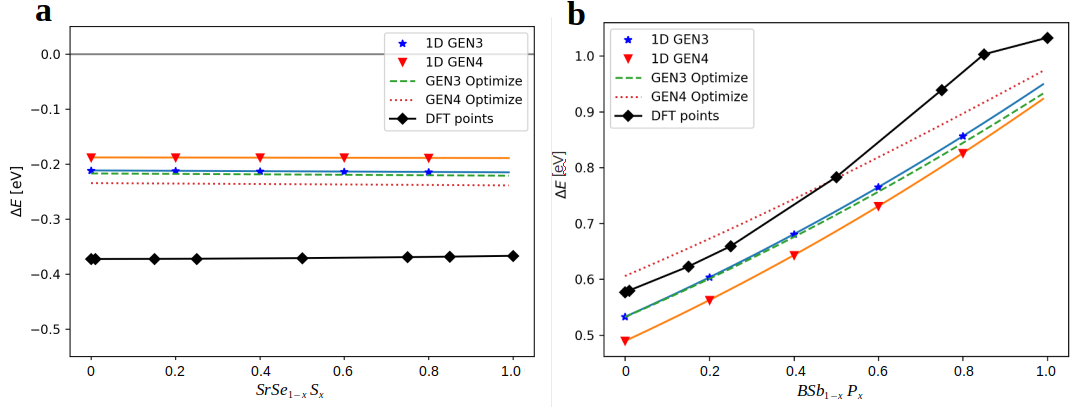}
    \caption{Total energy difference $\Delta E$ as a function of concentration ($x$) for a) $Sr[S_xSe_{1-x}]$ and b) $B[P_xSb_{1-x}]$ alloys, predicted from original and optimized 1D descriptors constructed using $GEN3$ and $GEN4$. Model predictions are verified using energy differences calculated via DFT\cite{DFT}\cite{DFT3}  (black-line with diamond points).}
    \label{fig:bsite1}
\end{figure*}

In the aim of further proving such trends and validate the implemented combinatorial ML method, we study the energetics in alloys of the type $[A_xA'_{1-x}]B$ and $A[B_xB'_{1-x}]$, where $x$ is the relative concentration of the mixing ions, monotonically tuning thus the average size of one ion with respect the other. All the alloy input properties  were linearly interpolated between corresponding values for end binaries ({\em i.e.} $AB$ and $A'B$ in the $[A_xA'_{1-x}]B$ case),  according to the Vegard's law \cite{Vegard}. 
For the A-ion mixing case, we considered SrSe, CaSe, MgSe, BeSe as parent AB compounds, already included in the original dataset. 
We then predicted the energy differences between RS and ZB phases for varying concentrations using the original and optimized 1D formulas constructed via $GEN3$ and $GEN4$ generators (Table~\ref{tab:1d} and Table~\ref{tab:1d_2}, respectively). To confirm the obtained predictions, we thus calculated the energy difference via DFT simulations, for a few intermediate concentrations.
The results, shown in Fig.~\ref{fig:a_site_change}, demonstrate an overall agreement between first-principles calculated and machine-learning predicted energetics. 
In particular, we notice a change of sign in  $\Delta E$, reflecting the change in the stability of the RS with respect to the ZB phase, when moving from the larger Strontium to the smaller Beryllium at the A-site, in line with the previously discussed relation between atomic radii of the A-ions and phase stabilization. 
At variance,  no such change of phase is observed when mixing ions at the B-site, keeping fixed the A-type one. This is confirmed, by looking at the energetics in $B[Sb_{1-x}P_x]$ and $Sr[Se_{1-x}S_x]$ alloys, shown in Fig.\ref{fig:bsite1}(a) and Fig.~\ref{fig:bsite1}(b), respectively. Despite the changing size of the average B-site, the two systems preserve the crystal structure adopted by the the parent compounds, {\em i.e.} rocksalt for the Sr-based compounds and zincblende for the B-based compounds. Such a behavior is still in line with preferred atomic structure fixed by the ion at the A-site, consistently with Strontium being larger than Boron. Qualitative agreement between ML-predicted and DFT-calculated energetics is observed again.
\\

After discussing the results related to 1D models, 
we now comment about the higher dimensional formulas. 
Our best 2D and 3D formulas from different generators are reported in Tables~\ref{tab:2d} and \ref{tab:3D}, respectively. 

\begin{figure*}
    \centering
      
    \includegraphics[scale=0.88]{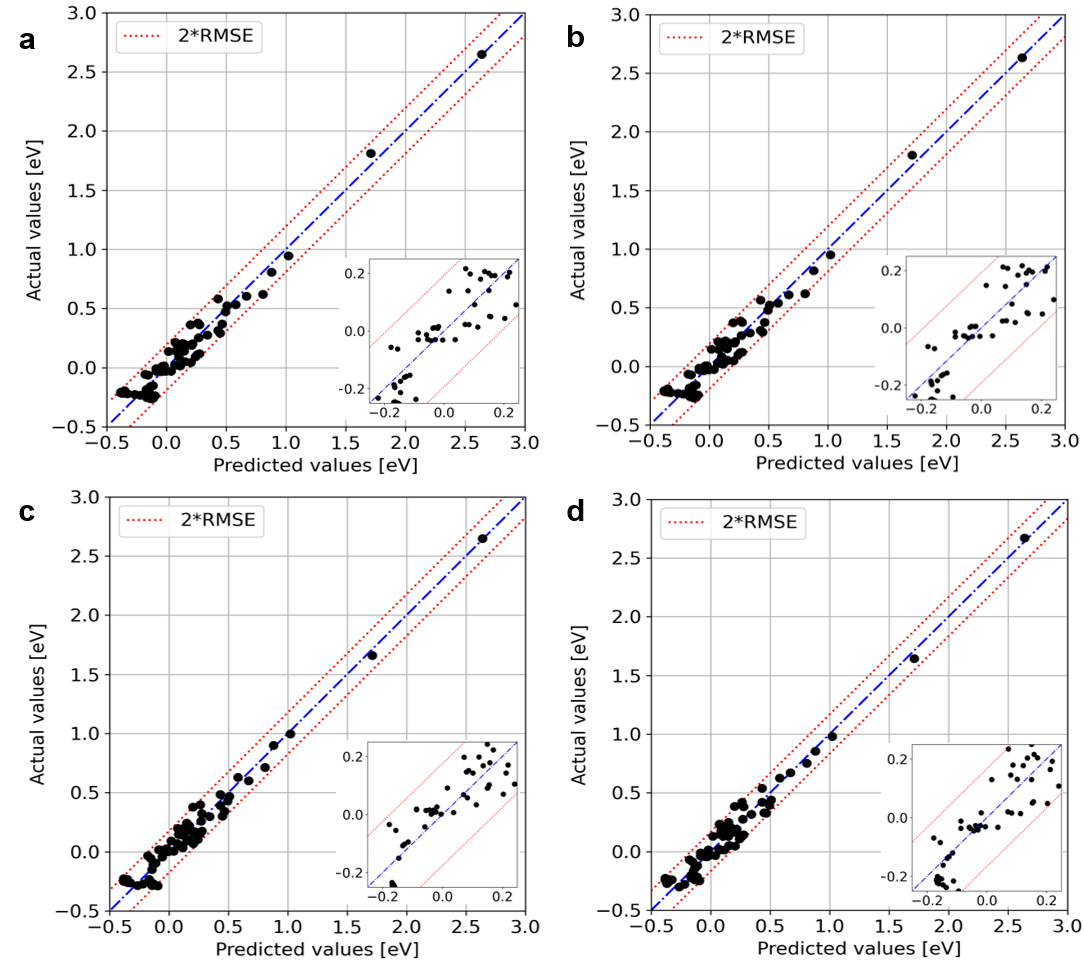}
    \caption{Comparison of actual ({\em i.e.} DFT) vs predicted total energy difference $\Delta E$ for a) 1D, c) 2D and  d) 3D  descriptors constructed using $GEN3$. Panel b) shows the best 1D descriptors after  formula optimization. Lower-right insets show a zoom in the relevant region where many compounds are concentrated. Red dotted lines correspond to 2$\times avg(RMSE)$ value. The  respective descriptors can be inferred from  tables-\ref{tab:1d}, \ref{tab:1d_2}, \ref{tab:2d}, \ref{tab:3D} }
    \label{fig:best_des}
\end{figure*}

To visualize the performance of the obtained formulas, we reproduce in Fig.~\ref{fig:best_des} the scatter plots of DFT-calculated energies as a function of model-predicted energy differences for the best formulas obtained by $GEN3$ - in terms of $avg(RMSE)$ - for 1D, 1D after formula optimization, 2D, and 3D models. From these, one can infer the quality of the prediction for the different approaches: 
the narrower the area between red lines (representing $ 2 \times avg(RMSE)$), the smaller the error or, equivalently, the more reliable the prediction. Notably, this is the case when building higher dimension formulas. 

In addition, a careful comparison between our results and those reported in the reference paper, Ref.\cite{ghiringhelli_big_2015}, is reported in Table S.1 of the Supplementary Material. In particular, in Fig. S.1 we compared the scatter plot of the 1D formula from $GEN3$ and Ref.\cite{ghiringhelli_big_2015},  with bar graphs of errors for individual compounds.
To check the improvement with respect to 1D formulas, we considered the $avg(RMSE)$ value, as also chosen in Ref.\cite{ghiringhelli_big_2015}. One can observe the improvement in $avg(RMSE)$ if we examine 1D and 2D formulas in Tables \ref{tab:1d} and \ref{tab:2d}. We notice around 10-20\% improvement from the original 1D to 2D, but less than 10\% of optimized 1D to original 2D formulas.
Furthermore, we also notice that original and optimized 1D formulas from $GEN3$ and $GEN4$ better perform with respect the corresponding 2D ones reported in Ref\cite{ghiringhelli_big_2015}. 

We remark that the process of formula optimization is less computationally expensive than the construction of higher-dimensional formulas. In addition, from the formula optimization one can gain a better physical insights about the contribution of individual primary atomic properties. 
These comments overall suggest that lower-dimensional formulas constitute a better choice in terms of physical interpretation and computational efficiency.


\section{ Conclusions}

The knowledge of a material stable crystal structure constitutes the starting point for any ab--initio modelling, since materials properties crucially depend on the periodic atomic arrangement in the crystal.  
Within this general framework, our aim here has been to exploit ML methods to
 correlate  the energetic stability of different crystal structures (zincblende vs rocksalt) for popular binary semiconducting compounds with primary properties of their atomic constituents, the latter representing simple and easily-accessible ingredients.  
 Based on atomic properties, 
we therefore built the material features using a combinatorial approach, we trained the machine learning model using the created features over a density--functional--theory dataset and we obtained simple mathematical expressions to quantitatively predict the energetic stability of one crystal structure over the other (i.e., a formula). In addition, we have also introduced an extra step following the linear regression to explore the relative contributions of individual basic atomic properties. 

To investigate the performance of the combinatorial approach, we compared our results with a reference paper \cite{ghiringhelli_big_2015}, where the authors predicted the stability of the crystal structure using an automated feature selection method. We found that our 1D formulas constructed using the combinatorial approach achieved a higher accuracy with respect to the reference ones. Furthermore, we  also learned more about the underlying mechanism from the formula optimization, where we found that the stability of RS and ZB heavily depends on the $r_p$ radius of A-sites. This kind of understanding is, in general, much more difficult to achieve in heavily-automated artificial--intelligence methods, such as neural networks, where it is not possible to interpret directly the model results. In this respect, our approach based on linear regression allows the construction of physical models supported by machine-driven suggestions of relevant ingredients; as such, it should be regarded as a methodology offering a huge range of applications in addressing microscopic mechanisms underlying different phenomena, calling for extensive investigations in the nearby future.

\section{AUTHOR DECLARATIONS}
\subsection{Conflict of Interest}
The authors have no conflicts to disclose.

\subsection{Data Availability}
The data that supports the findings of this study are available within the article and its supplementary material. The code for machine learning is available at https://github.com/lstorchi/
matinformatics .

\section{Supplementary material}
See Supplementary Material for technical details related to LR, DFT calculations of the alloy supercell, dataset and for additional results related to 1D, 2D, 3D formulas.

\section{ Acknowledgments} 
This project has received funding from the European Union’s Horizon 2020 research and innovation program under the Marie Sklodowska-Curie Grant Agreement No. 861145–BeMAGIC.
The authors acknowledge the Italian MIUR for supporting the PRIN project ``TWEET: Towards Ferroelectricity
in two dimensions”, grant n. 2017YCTB59, and the ``Nanoscience Foundries and Fine Analysis" (NFFA-MIUR Italy) project. 
Calculations were performed exploiting the computing resources at the Pharmacy Dept., Univ. Chieti-Pescara. 
D.A. is grateful to M. Verstraete (ULiege) for the time allowed to work on the writing of this paper. We are also thankful to L. Ghiringhelli for his fruitful support and insights.

\begin{table*}
  \centering
    
  \scalebox{1.}{
    \begin{tabular}{rlrrrrrr}
    \hline
    \multicolumn{1}{c}{\textbf{Formulas}} & \multicolumn{1}{c}{\textbf{avg(RMSE)}} & \multicolumn{1}{c}{\textbf{RMSE}} & \multicolumn{1}{c}{\textbf{$R^2$ }}&\multicolumn{1}{c}{Success rate} &
    \multicolumn{1}{c}{Generator type}  \\ \hline \hline

    \multicolumn{1}{c}{$0.117\cdot\frac{EA(B) - {IP(B)}}{r_p(A)^2}-0.342$} & \multicolumn{1}{c}{0.1455} & \multicolumn{1}{c}{0.1423} & \multicolumn{1}{c}{$0.89$} &
    \multicolumn{1}{c}{89\%} &\multicolumn{1}{c}{1D descriptor \cite{ghiringhelli_big_2015}}  \\ \hline

    \multicolumn{1}{c}{$ -0.751\cdot\frac{r_p(B)^{3} - exp[r_s(B)]}{r_p(A)^2}-0.317$} & \multicolumn{1}{c}{0.1296} & \multicolumn{1}{c}{0.1193} & \multicolumn{1}{c}{$0.92$} &
    \multicolumn{1}{c}{90\%} & \multicolumn{1}{c}{$GEN1$}    \\ \hline
    
    \multicolumn{1}{c}{$ 0.285\cdot\frac{\sqrt{|IP(B)} + \sqrt{|EA(A)|}}{r_p(A)^2}-0.387$ }
    & \multicolumn{1}{c}{0.1367} & \multicolumn{1}{c}{0.1309} & \multicolumn{1}{c}{0.91} & \multicolumn{1}{c}{91\%} & \multicolumn{1}{c}{$GEN2$}    \\ \hline
    
    \multicolumn{1}{c}{$0.774\cdot\frac{r_p(B) + \sqrt{|r_d(A)|}}{r_p(A)^3 + r_p(B)^3} -0.303 $} &  
    \multicolumn{1}{c}{0.0995} & \multicolumn{1}{c}{0.0963} & 
    \multicolumn{1}{c}{0.95} & \multicolumn{1}{c}{94\%} & 
    \multicolumn{1}{c}{$GEN3$}    \\ \hline
    
    \multicolumn{1}{c}{$ 1.155\cdot\frac{r_s(B) + r_s(A)}{r_p(B)^3 + r_p(A)^3}-0.368$} &  
    \multicolumn{1}{c}{0.1103} &  \multicolumn{1}{c}{0.1058} & 
    \multicolumn{1}{c}{0.94} & 
    \multicolumn{1}{c}{96\%} &
    \multicolumn{1}{c}{$GEN4$}    \\ 
    \hline
    
    \end{tabular}}
    \caption{1D formulas, along with  related statistics: $avg(RMSE)$ denotes the root mean squared error for average over 1000 random train-test splits of dataset. Instead, the RMSE is the root mean squared error for the entire dataset as training and test. Similarly, the  $R^2$ values are calculated considering the entire dataset and they show the quality of fit between predicted and actual values. The success rate (in percent) shows how many RS or ZB phases out of 82 have been  correctly identified by the descriptor. The ``Generator type" column indicates the different generators used to produce the corresponding descriptor. RMSEs are in eV.}
    \label{tab:1d}
    \hspace{5em}
    
    \scalebox{1.}{
    \begin{tabular}{rlrrrrrrrr}
    \hline
    
    \multicolumn{1}{c}{\textbf{Formula}} & \multicolumn{1}{c}{\textbf{avg(RMSE)}} & \multicolumn{1}{c}{\textbf{RMSE}} &
    \multicolumn{1}{c}{\textbf{$R^2$}} & \multicolumn{1}{c}{\textbf{Success Rate}} &
    \multicolumn{1}{c}{Generator type} \\ \hline \hline

    \multicolumn{1}{c}{$ 0.127\cdot\frac{0.800\cdot EA(B) - 1.000\cdot {IP(B)}}{1.110\cdot r_p(A)^2}-0.352$} & \multicolumn{1}{c}{0.1457} & \multicolumn{1}{c}{0.1419} & 0.89 & \multicolumn{1}{c}{89\%}& \multicolumn{1}{c}{1D descriptor \cite{ghiringhelli_big_2015}}    \\ \hline

    \multicolumn{1}{c}{$ -1.870\frac{0.801\cdot \sqrt{r_p(B)} - 0.606\cdot exp[r_p(A)]}{1.010\cdot r_p(A)^3}-0.968$} & \multicolumn{1}{c}{0.1191} &  \multicolumn{1}{c}{0.1143} & 0.93&\multicolumn{1}{c}{91\%} & \multicolumn{1}{c}{$GEN1$}    \\ \hline
    
    \multicolumn{1}{c}{$ 0.477\cdot\frac{0.876\cdot \sqrt{|HOMO(B)|} + 0.468\cdot \sqrt{|LUMO(B)|}}{1.110\cdot r_p(A)^2}-0.372 $ }
    & \multicolumn{1}{c}{0.1340} & \multicolumn{1}{c}{0.1296} &0.91 &\multicolumn{1}{c}{91\%}& \multicolumn{1}{c}{$GEN2$}    \\ \hline
    
    \multicolumn{1}{c}{$ 1.609\cdot \frac{0.642\cdot r_p(B) + 0.502\cdot \sqrt{|r_d(A)|}}{1.170\cdot r_p(A)^3 + 1.170\cdot r_p(B)^3}-0.309$} &  
    \multicolumn{1}{c}{0.0991} & 
    \multicolumn{1}{c}{0.0961} & 0.95&\multicolumn{1}{c}{94\%} &
    \multicolumn{1}{c}{$GEN3$}    \\ \hline

    \multicolumn{1}{c}{$ 1.207\cdot\frac{0.878\cdot r_s(B) + 0.200\cdot r_p(A)}{0.512\cdot r_p(B)^3 + 0.610\cdot r_p(A)^3}-0.359$} &  
    \multicolumn{1}{c}{0.1045} & \multicolumn{1}{c}{0.1016} & 0.94 & \multicolumn{1}{c}{99\%}&
    \multicolumn{1}{c}{$GEN4$}    \\ 
    \hline
    
    \end{tabular}}
    \caption{1D formulas after the optimization step, along with related statistics. Notation as in table-\ref{tab:1d}.}
    \label{tab:1d_2}
  
\end{table*}%

\begin{table*}
  \centering
    
  \scalebox{1}{
    \begin{tabular}{rlrrrrrr}
    \hline
    \multicolumn{1}{c}{\textbf{Descriptor}} & \multicolumn{1}{c}{\textbf{avg(RMSE)}} & \multicolumn{1}{c}{\textbf{RMSE}} &
    \multicolumn{1}{c}{\textbf{$R^2$}} & \multicolumn{1}{c}{\textbf{Success Rate}} &
    \multicolumn{1}{c}{Generator type} \\ \hline \hline

    \multicolumn{1}{c}{$ 0.113\cdot\frac{{EA(B)} - {IP(B)}}{r_p(A)^2}-1.558\cdot\frac{|r_s(A)-r_p(B)|}{exp[r_s(A)]} -0.133 $} & \multicolumn{1}{c}{0.1041} & \multicolumn{1}{c}{0.0988} & \multicolumn{1}{c}{0.95} & \multicolumn{1}{c}{96\%} & \multicolumn{1}{c}{2D descriptor \cite{ghiringhelli_big_2015}}    \\ \hline
    
    \multicolumn{1}{c}{$ -0.342\cdot\frac{{r_p(B)}^3 - {exp[r_p(A)]}}{r_p(A)^3} -1.042\cdot\frac{r_p(A)^2-\sqrt{|r_d(A)|}}{exp[r_p(A)]} -0.062$} & \multicolumn{1}{c}{0.0989} & \multicolumn{1}{c}{0.0944} & \multicolumn{1}{c}{0.95} & \multicolumn{1}{c}{89\%} & \multicolumn{1}{c}{$GEN1$}    \\  \hline
    
    \multicolumn{1}{c}{$-0.081\cdot\frac{IP(B) + \sqrt{|IP(A)|}}{r_p(A)^3}-0.001\cdot\frac{r_s(A)^3 - \sqrt{r_d(A)}}{exp(HOMOKS(A))} -0.062$} & 
    \multicolumn{1}{c}{0.1163} & \multicolumn{1}{c}{0.1100} & \multicolumn{1}{c}{0.93} & \multicolumn{1}{c}{86\%} &
    \multicolumn{1}{c}{$GEN2$}   \\  \hline
    
    \multicolumn{1}{c}{
    $-1.175\cdot\frac{r_p(A) - \sqrt{|r_d(A)|}}{r_s(B)^3 + r_p(A)^3} +  0.513\cdot\frac{r_s(B) + \sqrt{|r_p(B)|}}{r_p(B)^3 + r_s(A)^3} - 0.250$ }& 
    \multicolumn{1}{c}{0.0911} & \multicolumn{1}{c}{0.0878} & \multicolumn{1}{c}{0.96} & \multicolumn{1}{c}{87\%} & \multicolumn{1}{c}{$GEN3$}  \\  \hline
    
    \multicolumn{1}{c}{
    $ 0.618\cdot\frac{r_d(A)/ {r_p(B)}}{r_p(A)^3 * \sqrt{r_d(A)}} + 
      1.097\cdot\frac{r_p(A) * \sqrt{|r_p(B)|}}{r_p(B)^3 + r_p(A)^3}-0.384 $ }& 
    \multicolumn{1}{c}{0.0995} & \multicolumn{1}{c}{0.0955} & \multicolumn{1}{c}{0.95} &
    \multicolumn{1}{c}{92\%} &\multicolumn{1}{c}{$GEN4$}  \\  \hline

    \end{tabular}}
    \caption{2D descriptors, along with related statistics. Notation as in table-\ref{tab:1d}.}
  \label{tab:2d}
  
\end{table*}

\begin{table*}
  \centering
    
  \scalebox{.8}{
    \begin{tabular}{rlrrrrrr}
    \hline
    \multicolumn{1}{c}{\textbf{Descriptor}} & \multicolumn{1}{c}{\textbf{avg(RMSE)}} & \multicolumn{1}{c}{\textbf{RMSE}} &
    \multicolumn{1}{c}{\textbf{$R^2$}} & \multicolumn{1}{c}{\textbf{Success Rate}} &
    \multicolumn{1}{c}{Generator type} \\ \hline \hline
    
    \multicolumn{1}{c}
   {
   $0.108\cdot\frac{{EA(B)} - {IP(B)}}{r_p(A)^2} -1.806\cdot 
   \frac{|r_s(A)-r_p(B)|}{exp[r_s(A)]}-3.782\cdot
   \frac{|r_p(B)-r_s(B)|}{exp[r_d(A)]} -0.023$ }
   & \multicolumn{1}{c}{0.0818} & \multicolumn{1}{c}{0.0756} & \multicolumn{1}{c}{0.97} & \multicolumn{1}{c}{93\%} & \multicolumn{1}{c}{3D descriptor \cite{ghiringhelli_big_2015}}    \\ \hline

   \multicolumn{1}{c}
   {  
   $0556\cdot\frac{r_p(B)^3- exp[r_p(A)]}{r_p(A)^3}+0.364\cdot 
   \frac{r_p(A)^2-\sqrt{|r_d(B)|}}{exp[r_p(A)]},
   -0.124\cdot\frac{r_p(B)^2-\sqrt{|r_d(A)|}}{r_p(A)^3}-1.87$ }
   & \multicolumn{1}{c}{0.1003} & \multicolumn{1}{c}{0.0933} & \multicolumn{1}{c}{0.95} & \multicolumn{1}{c}{90\%}& \multicolumn{1}{c}{$GEN1$}    \\  \hline
    
    \multicolumn{1}{c}
    {  
    $  -0.056 \cdot\frac{(LUMOKS(A) + HOMOKS(B))}{r_p(A)^3}  
    +0.266\cdot\frac{\sqrt{|EA(B)|} + exp(EA(B))}{r_s(A)^3} 
     -0.016\cdot\frac{HOMOKS(A) - exp(LUMOKS(B))}{(r_p(A)^3)}-0.310   $ } 
    & \multicolumn{1}{c}{0.1300}  & \multicolumn{1}{c}{0.1205} & \multicolumn{1}{c}{0.92} & \multicolumn{1}{c}{91\%}&  \multicolumn{1}{c}{ $GEN2$}    \\ \hline

   \multicolumn{1}{c}
   {  
   $ -0.885\cdot\frac{r_p(B) - exp[r_p(A)]}{r_p(A)^2 + r_p(A)^3}    
    -0.417\cdot\frac{r_s(A) - exp[r_s(B)]}{r_s(A)^3 + r_p(B)^3} 
   -0.579\cdot\frac{r_p(A) - \sqrt{|r_d(A)|}}{r_p(B)^2 + r_s(A)^3}-0.616$ } 
   & \multicolumn{1}{c}{0.0875} & \multicolumn{1}{c}{0.0834} & \multicolumn{1}{c}{0.96} & \multicolumn{1}{c}{98\%} & \multicolumn{1}{c}{$GEN3$}    \\ \hline
   
     \multicolumn{1}{c}
   {  
   $ 0.635\cdot\frac{\sqrt{IP(B)}/ \sqrt{IP(A)}]}{r_p(A)^3 + r_p(B)^3}$     
   $ +0.730\cdot\frac{r_p(B) * \sqrt{|r_d(A)|}}{r_p(A)^3 + r_p(B)^3}$ 
   $+0.038\cdot\frac{IP(A)^2 - EA(A)^2}{exp(r_p(A)) * exp(r_d(B))}-0.358$ } 
   & \multicolumn{1}{c}{0.0989} & \multicolumn{1}{c}{0.0919} & \multicolumn{1}{c}{0.96} &  \multicolumn{1}{c}{93\%} & \multicolumn{1}{c}{$GEN4$}    \\ \hline
    \end{tabular}}
    \caption{3D descriptors, along with related statistics. Notation as in table-\ref{tab:1d}.}
    \label{tab:3D}
\end{table*}

\clearpage
\bibliography{RS_ZB_ref.bib}

\end{document}


{\centering \bf \Large Supplementary Material: \\``Towards machine learning for microscopic mechanisms: a formula search for crystal structure stability based on atomic properties''}

\vspace{1cm}

{\bf Udaykumar Gajera$^{1,2}$, Loriano Storchi$^3$, Danila Amoroso$^{1,4}$, Francesco Delodovici$^1$, Silvia Picozzi$^1$}

1. Consiglio Nazionale delle Ricerche, CNR-SPIN c/o Università ``G. D’Annunzio'', 66100 Chieti, Italy 

2. Department of Chemical and Material Science, Università degli studi di Torino, 10124 Torino, Italy 

3. Dipartimento di Farmacia, Universitá degli Studi ``G. D’Annunzio'', 66100 Chieti, Italy

4. NanoMat/Q-mat/CESAM,Universite de Liege, B-4000 Liege, Belgium

\section{Validation of the linear regression}\label{varification}
We employed different verification parameters to check the model's efficiency, namely: RMSE, Pearson correlation coefficient, $R^2$ values, and classification accuracy.  RMSE represents the root mean square error of the test dataset. 
The Pearson correlation coefficient, defined in equation \ref{eq_pear}, represents a measurement of input and output property dependence \cite{ref1}. If two properties are highly dependent one on the other, one gets values closer to 1 or -1, whereas values closer to zero show a much lower dependence.   $R^2$, {\em i.e.} the coefficient of determination defined in equation \ref{eq_rsqua}, describes how properly the regression line interpolates the data. Finally, the classification accuracy shows the ability of the linear model to qualitatively distinguish different classes of the dataset, in our case RS and ZB as stable phases. 
\begin{equation}
\label{eq_pear}
    r = \frac{\sum (x_i- \Bar{x})(y_i-\Bar{y})}{\sqrt{\sum(x_i- \Bar{x})^2\sum(y_i-\Bar{y})^2 }}
\end{equation}
\vspace{2em}
\begin{equation}
\label{eq_rsqua}
    R^2 = 1- \frac{SS_{residual}}{SS_{total}}
\end{equation}
In equation \ref{eq_rsqua}: $SS_{total} = \sum_i(y_i - \Bar{y})^2$, $SS_{residual} = \sum_i(y_i - f_i)^2$ where $f_i$ are the predicted values; $x_i$ and $y_i$ are values of input and output properties; $\Bar{x}$ and $\Bar{y}$ are mean of the input and output values.

\section{Analysis of the best 1D, 2D and 3D descriptors}

Table-\ref{tab:my_label} reports other verification parameters calculated for the best descriptors of each generator, including those presented in the paper. The relevance of calculating avg(RMSE train) and avg(RMSE test) lies in analysing the bias-variance tradeoff\cite{Rajnarayan2010} in LR.  $\rm Max\_E$ and $\rm Min\_E$ indicate the maximum and minimum error in the prediction for the specific descriptor.

\begin{figure}
 \renewcommand\thefigure{S.1}
    \centering
    \captionsetup{justification=justified,singlelinecheck=false}
    \includegraphics[scale=0.86]{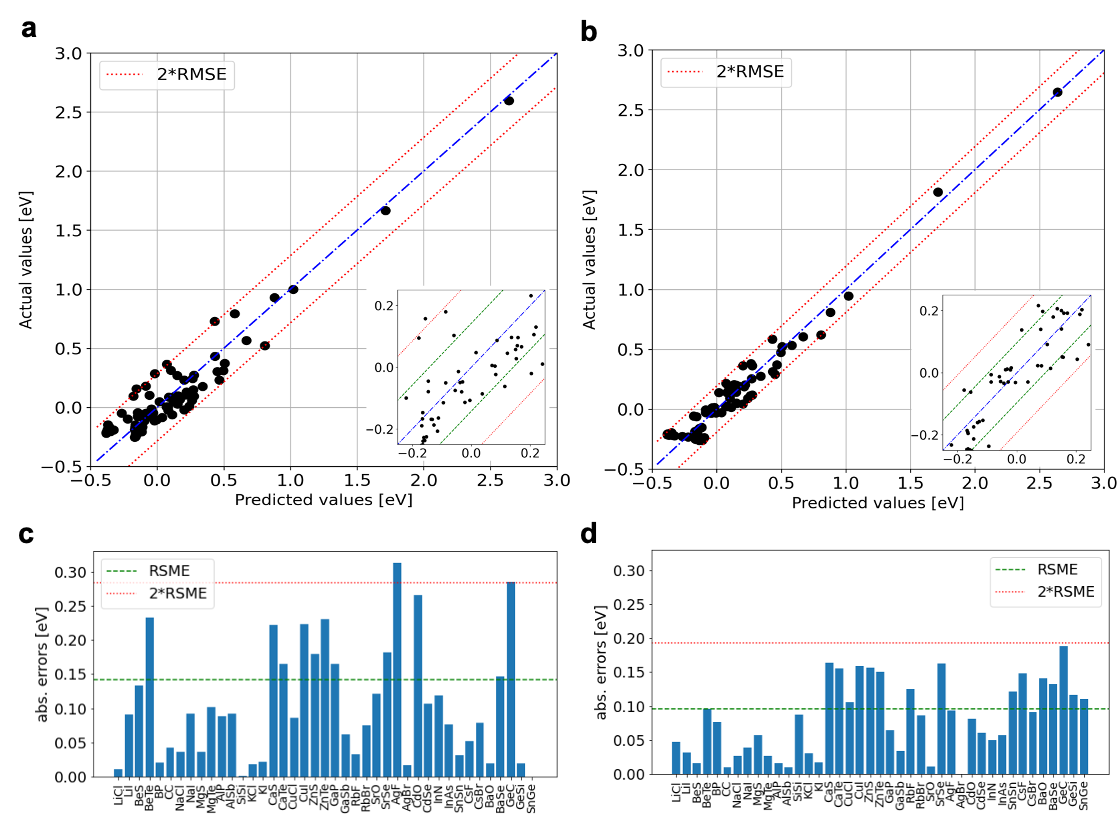}
    \caption{Panel \textbf{a} reports the predicted against actual ({\em i.e} DFT) $\Delta E$ values for 1D descriptor presented in Ref. \cite{ghiringhelli_big_2015}; panel \textbf{b} reports those obtained by $GEN3$ as a comparison. Absolute error for the same formula for different compounds in the bar graph (panels \textbf{c} and \textbf{d}). The related descriptors used to calculate the values can be inferred from the main text.}
    \label{fig:my_label1}
\end{figure}

Figure \ref{fig:my_label1} reports in panel \textbf{a}(\textbf{b}) the {\it ab-initio} energies against the energies predicted through the 1D descriptor proposed in Ref.\cite{ghiringhelli_big_2015} (obtained within GEN3). In addition,  panels \textbf{(c)} and \textbf{(d)} of figure \ref{fig:my_label1} report the absolute errors obtained employing the 1D descriptors mentioned above, for certain compounds. From a comparison of the two scatter plots and of the two bar-graphs, one can infer the improved accuracy obtained with the GEN3 descriptor.

\begin{table}
 \renewcommand\thetable{S.1}
\captionsetup{justification=justified,singlelinecheck=false}
\scalebox{0.72}{\centering

\begin{tabular}{ccccccccccc}
\toprule
{} &  Details &  avg(RSME train) &  avg(RMSE test) &  RMSE &  $R^2$ &  Pearson\_coeff & success rate &     Max\_E &     Min\_E &  Std\_daviation \\
\midrule
0  &   Ref\_1D\cite{ghiringhelli_big_2015} &        0.1420 &       0.1455 &   0.1422 &     0.89 &          0.947 &       89\% &  0.0523 &  0.0041 &       0.0081 \\
1  &  1D\_GEN1 &        0.1186 &       0.1296 &   0.1192 &     0.92 &          0.963 &       90\% &  0.0743 &  0.0014 &       0.0083 \\
2  &  1D\_GEN2 &        0.1305 &       0.1367 &   0.1309 &     0.91 &          0.956 &       91\% &  0.0676 &  0.0027 &       0.0105 \\
3  &  1D\_GEN3 &        0.0961 &       0.0995 &   0.0963 &     0.95 &          0.976 &        94\% &  0.0234 &  0.0016 &       0.0032 \\
4  &  1D\_GEN4 &        0.1055 &       0.1103 &   0.1058 &     0.94 &          0.971 &       96\% &  0.0330 &  0.0012 &       0.0044 \\
5  &   Ref\_2d\cite{ghiringhelli_big_2015} &        0.0983 &       0.1041 &   0.0987 &     0.95 &          0.975 &       96\% &  0.0323 &  0.0019 &       0.0044 \\
6  &  2D\_GEN1 &        0.0941 &       0.0988 &   0.0943 &     0.95 &          0.977 &       89\% &  0.0419 &  0.0020 &       0.0040 \\
7  &  2D\_GEN2 &        0.1095 &       0.1163 &   0.1099 &     0.93 &          0.969 &       87\% &  0.0489 &  0.0011 &       0.0083 \\
8  &  2D\_GEN3 &        0.0875 &       0.0911 &   0.0878 &     0.96 &          0.980 &        88\% &  0.0178 &  0.0014 &       0.0026 \\
9  &  2D\_GEN4 &        0.0951 &       0.0995 &   0.0954 &     0.95 &          0.977 &       93\% &  0.0221 &  0.0016 &       0.0033 \\
10 &   Ref\_3d\cite{ghiringhelli_big_2015} &        0.0751 &       0.0814 &   0.0755 &     0.97 &          0.985 &       93\% &  0.0185 &  0.0009 &       0.0031 \\
11 &  3D\_GEN1 &        0.0929 &       0.1003 &   0.0933 &     0.95 &          0.978 &       90\% &  0.0282 &  0.0024 &       0.0038 \\
12 &  3D\_GEN2 &        0.1200 &       0.1300 &   0.1205 &     0.92 &          0.963 &       91\% &  0.1119 &  0.0021 &       0.0103 \\
13 &  3D\_GEN3 &        0.0832 &       0.0874 &   0.0834 &     0.96 &          0.982 &       98\% &  0.0199 &  0.0015 &       0.0026 \\
14 &  3D\_GEN4 &        0.0915 &       0.0989 &   0.0919 &     0.96 &          0.978 &       93\% &  0.0227 &  0.0013 &       0.0035 \\
\bottomrule

\end{tabular}}
\caption{Different verification parameters for 1D, 2D and 3D descriptors calculated in the present work and descriptors presented in Ref.\cite{ghiringhelli_big_2015}. Here, avg(RMSE train), avg(RMSE test) and RMSE  indicate the root mean squared error for training data, test data and full dataset, respectively. $R^2$ and Pearson coeff are goodness parameters.  Max\_E and Min\_E show the maximum and minimum absolute error in prediction.}
    \label{tab:my_label}
\end{table}

\subsection{Dependence of energy difference on the atomic features}
\begin{figure}
 \renewcommand\thefigure{S.2}
    \centering
    \captionsetup{justification=justified,singlelinecheck=false}
    \includegraphics[scale=0.6]{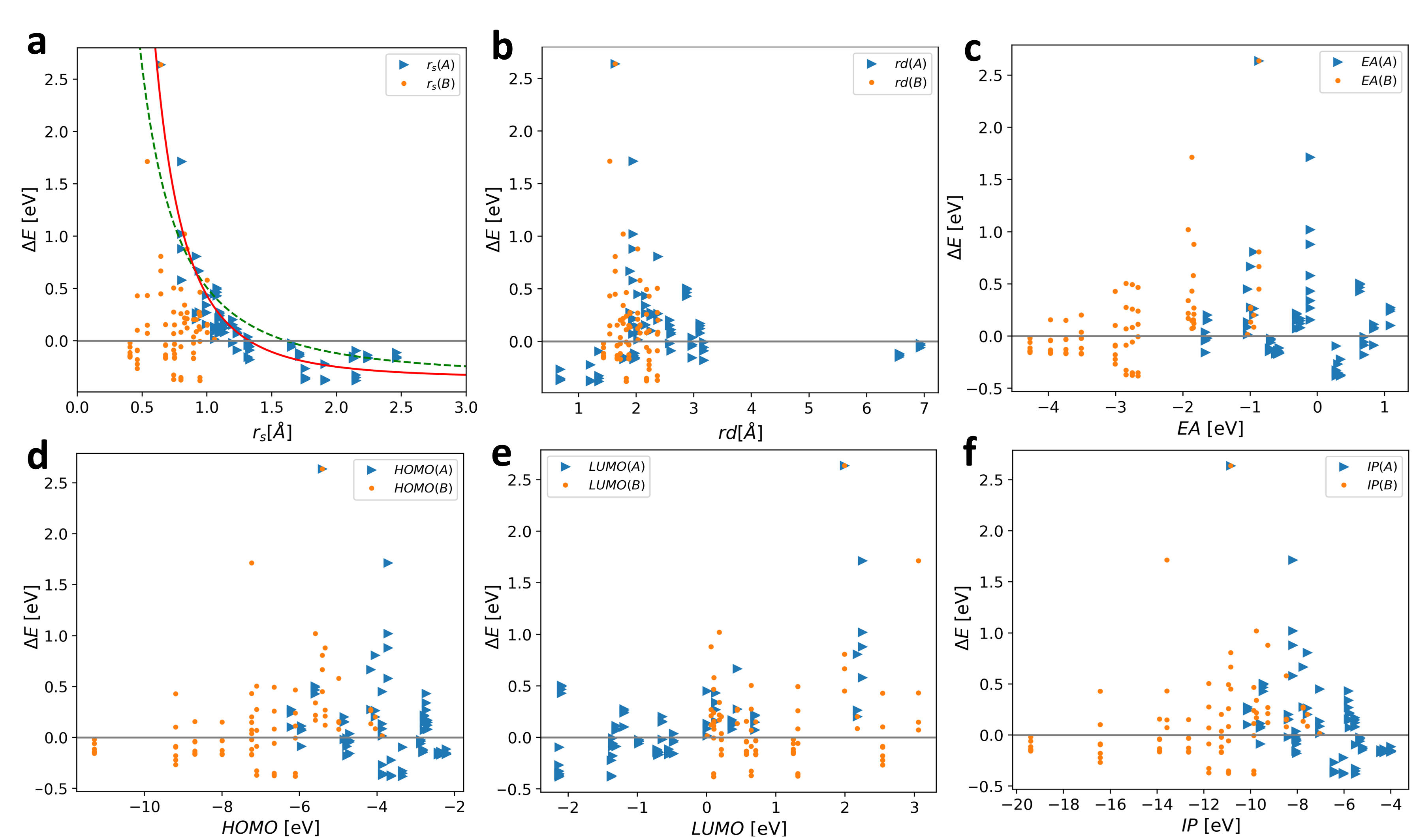}
    \caption{Dependence of $\Delta E$ on atomic features: a) $r_s$, b) $r_d$, c) $EA$, d) $LUMO$, e) $HOMO$ and f) $IP$. Orange dots (blue triangles) indicate values relative to the A (B) atoms. In panel-\textbf{a}, we perform a fit using a function $f(x)$   proportional to $x^{-2}$ (dotted green line) and to $x^{-3}$ (straight red line).} 
    \label{fig:basic_vs_de}
\end{figure}

Figure \ref{fig:basic_vs_de} reports the dependence of the total energy difference between  RS and ZB phases as a function of the atomic features, except for r$_p$ (since that is reported in the main text). It appears clearly that  only $r_s$ is strongly correlated with the energy difference, at variance with other atomic features where the correlation is small or absent.

\subsection{Formula optimization using automated optimization methods}
\begin{figure}
 \renewcommand\thefigure{S.3}

    \centering
    \captionsetup{justification=justified,singlelinecheck=false}
    \includegraphics[scale=0.66]{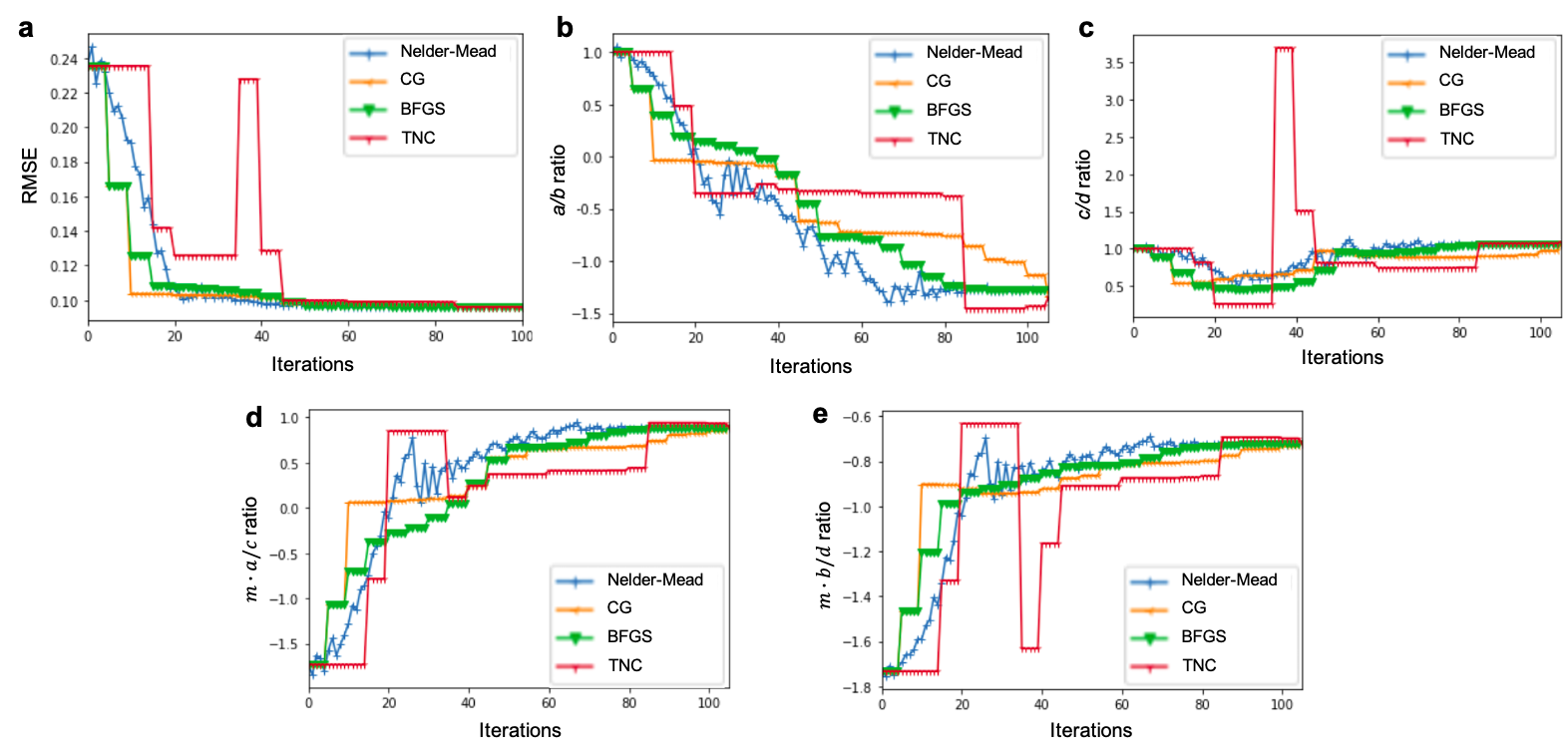}
    \caption{Evolution of different parameters at each iteration using different automated optimizing algorithms: Nelder-Mead (Blue lines), CG (orange lines), BFGS (green lines) and TNC (red lines). Here, we show the evolution of RSME, $a/b, c/d, m\times a/c$ and $m \times b/d$ in panels  \textbf{a, b, c, d} and \textbf{e} respectively.}
    \label{fig:automated_methods}
\end{figure}

To find the relative contribution of individual atomic features in the descriptor, we employed a grid search method. We further bench-marked our grid-search optimization for the best 1D descriptor constructed using GEN3, defined in equation \ref{opt_gen3}, by means of other automated optimization methods: Nelder-Mead\cite{doi:10.1137/1.9781611973501.ch18}, Conjugate Gradient(CG)\cite{golub_matrix_2013}, Broyden–Fletcher–Goldfarb–Shanno(BFGS)\cite{doi:10.1137/1.9781611970920.ch4} and truncated Newton(TNC)\cite{dembo_inexact_1982}: 

\begin{equation}
    \frac{r_p(B) + \sqrt{|r_d(A)|}}{r_p(A)^3 + r_p(B)^3} 
    \label{opt_gen3}
\end{equation}
if we introduce different coefficients (a,b,c,d) multiplying each atomic feature, then the expression can be written as:
\begin{equation}
    \frac{a\cdot r_p(B) + b\cdot\sqrt{|r_d(A)|}}{c\cdot r_p(A)^3 +d\cdot r_p(B)^3} 
\end{equation}
Through the automated optimization methods mentioned above, one can obtain the set of coefficients that give the lowest RMSE. Each step of the optimization is followed by LR. Thus, the optimization modifies the slope (m) and intercept moving towards the coefficients combination with the lowest possible RMSE. From 80 to 100 iterations are needed for each method to reach the minimum RMSE, as reported in panel \textbf{a} of figure \ref{fig:automated_methods}.
Panels from \textbf{b} to \textbf{e} report the trend of the ratios $a/b, c/d, m \times a/c$ and $m \times b/d$. All these quantities appear to converge to constant values in the final steps of the optimization.

\begin{table}[h]
 \renewcommand\thetable{S.2}
    \centering
    \resizebox{\columnwidth}{!}{%
    \begin{tabular}{lllrrrrrrrrrrrrrrr}
\toprule A &B & \multicolumn{1}{p{2cm}}{\centering DFT \\ Classification} &     $\Delta E$ &
  IP(A) &   EA(A) &  HOMO(A) &  LUMO(A) &   $r_s$(A) &   $r_p$(A) &   $r_d$(A) &    IP(B) &   EB(B) &  HOMO(B) &  LUMO(B) &   $r_s$(B) &   $r_p$(B) &   $r_d$(B) \\
\midrule
 Li &   F &        RS & -0.059 &  -5.329 & -0.698 &  -2.874 &  -0.978 &  1.652 &  1.995 &  6.930 & -19.404 & -4.273 & -11.294 &   1.251 &  0.406 &  0.371 &  1.428 \\
 Li &  Cl &        RS & -0.038 &  -5.329 & -0.698 &  -2.874 &  -0.978 &  1.652 &  1.995 &  6.930 & -13.902 & -3.971 &  -8.700 &   0.574 &  0.679 &  0.756 &  1.666 \\
 Li &  Br &        RS & -0.033 &  -5.329 & -0.698 &  -2.874 &  -0.978 &  1.652 &  1.995 &  6.930 & -12.650 & -3.739 &  -8.001 &   0.708 &  0.749 &  0.882 &  1.869 \\
 Li &   I &        RS & -0.022 &  -5.329 & -0.698 &  -2.874 &  -0.978 &  1.652 &  1.995 &  6.930 & -11.257 & -3.513 &  -7.236 &   0.213 &  0.896 &  1.071 &  1.722 \\
 Be &   O &        ZB &  0.430 &  -9.459 &  0.631 &  -5.600 &  -2.098 &  1.078 &  1.211 &  2.877 & -16.433 & -3.006 &  -9.197 &   2.541 &  0.462 &  0.427 &  2.219 \\
 Be &   S &        ZB &  0.506 &  -9.459 &  0.631 &  -5.600 &  -2.098 &  1.078 &  1.211 &  2.877 & -11.795 & -2.845 &  -7.106 &   0.642 &  0.742 &  0.847 &  2.366 \\
 Be &  Se &        ZB &  0.495 &  -9.459 &  0.631 &  -5.600 &  -2.098 &  1.078 &  1.211 &  2.877 & -10.946 & -2.751 &  -6.654 &   1.316 &  0.798 &  0.952 &  2.177 \\
 Be &  Te &        ZB &  0.466 &  -9.459 &  0.631 &  -5.600 &  -2.098 &  1.078 &  1.211 &  2.877 &  -9.867 & -2.666 &  -6.109 &   0.099 &  0.945 &  1.141 &  1.827 \\
  B &   N &        ZB &  1.713 &  -8.190 & -0.107 &  -3.715 &   2.248 &  0.805 &  0.826 &  1.946 & -13.585 & -1.867 &  -7.239 &   3.057 &  0.539 &  0.511 &  1.540 \\
  B &   P &        ZB &  1.020 &  -8.190 & -0.107 &  -3.715 &   2.248 &  0.805 &  0.826 &  1.946 &  -9.751 & -1.920 &  -5.596 &   0.183 &  0.826 &  0.966 &  1.771 \\
  B &  As &        ZB &  0.879 &  -8.190 & -0.107 &  -3.715 &   2.248 &  0.805 &  0.826 &  1.946 &  -9.262 & -1.839 &  -5.341 &   0.064 &  0.847 &  1.043 &  2.023 \\
  C &   C &        ZB &  2.638 & -10.852 & -0.872 &  -5.416 &   1.992 &  0.644 &  0.630 &  1.631 & -10.852 & -0.872 &  -5.416 &   1.992 &  0.644 &  0.630 &  1.631 \\
 Na &   F &        RS & -0.146 &  -5.223 & -0.716 &  -2.819 &  -0.718 &  1.715 &  2.597 &  6.566 & -19.404 & -4.273 & -11.294 &   1.251 &  0.406 &  0.371 &  1.428 \\
 Na &  Cl &        RS & -0.133 &  -5.223 & -0.716 &  -2.819 &  -0.718 &  1.715 &  2.597 &  6.566 & -13.902 & -3.971 &  -8.700 &   0.574 &  0.679 &  0.756 &  1.666 \\
 Na &  Br &        RS & -0.127 &  -5.223 & -0.716 &  -2.819 &  -0.718 &  1.715 &  2.597 &  6.566 & -12.650 & -3.739 &  -8.001 &   0.708 &  0.749 &  0.882 &  1.869 \\
 Na &   I &        RS & -0.115 &  -5.223 & -0.716 &  -2.819 &  -0.718 &  1.715 &  2.597 &  6.566 & -11.257 & -3.513 &  -7.236 &   0.213 &  0.896 &  1.071 &  1.722 \\
 Mg &   O &        RS & -0.178 &  -8.037 &  0.693 &  -4.782 &  -1.358 &  1.330 &  1.897 &  3.171 & -16.433 & -3.006 &  -9.197 &   2.541 &  0.462 &  0.427 &  2.219 \\
 Mg &   S &        RS & -0.087 &  -8.037 &  0.693 &  -4.782 &  -1.358 &  1.330 &  1.897 &  3.171 & -11.795 & -2.845 &  -7.106 &   0.642 &  0.742 &  0.847 &  2.366 \\
 Mg &  Se &        RS & -0.055 &  -8.037 &  0.693 &  -4.782 &  -1.358 &  1.330 &  1.897 &  3.171 & -10.946 & -2.751 &  -6.654 &   1.316 &  0.798 &  0.952 &  2.177 \\
 Mg &  Te &        RS & -0.005 &  -8.037 &  0.693 &  -4.782 &  -1.358 &  1.330 &  1.897 &  3.171 &  -9.867 & -2.666 &  -6.109 &   0.099 &  0.945 &  1.141 &  1.827 \\
 Al &   N &        ZB &  0.072 &  -5.780 & -0.313 &  -2.784 &   0.695 &  1.092 &  1.393 &  1.939 & -13.585 & -1.867 &  -7.239 &   3.057 &  0.539 &  0.511 &  1.540 \\
 Al &   P &        ZB &  0.219 &  -5.780 & -0.313 &  -2.784 &   0.695 &  1.092 &  1.393 &  1.939 &  -9.751 & -1.920 &  -5.596 &   0.183 &  0.826 &  0.966 &  1.771 \\
 Al &  As &        ZB &  0.212 &  -5.780 & -0.313 &  -2.784 &   0.695 &  1.092 &  1.393 &  1.939 &  -9.262 & -1.839 &  -5.341 &   0.064 &  0.847 &  1.043 &  2.023 \\
 Al &  Sb &        ZB &  0.150 &  -5.780 & -0.313 &  -2.784 &   0.695 &  1.092 &  1.393 &  1.939 &  -8.468 & -1.847 &  -4.991 &   0.105 &  1.001 &  1.232 &  2.065 \\
 Si &   C &        ZB &  0.668 &  -7.758 & -0.993 &  -4.163 &   0.440 &  0.938 &  1.134 &  1.890 & -10.852 & -0.872 &  -5.416 &   1.992 &  0.644 &  0.630 &  1.631 \\
 Si &  Si &        ZB &  0.275 &  -7.758 & -0.993 &  -4.163 &   0.440 &  0.938 &  1.134 &  1.890 &  -7.758 & -0.993 &  -4.163 &   0.440 &  0.938 &  1.134 &  1.890 \\
  K &   F &        RS & -0.146 &  -4.433 & -0.621 &  -2.426 &  -0.697 &  2.128 &  2.443 &  1.785 & -19.404 & -4.273 & -11.294 &   1.251 &  0.406 &  0.371 &  1.428 \\
  K &  Cl &        RS & -0.165 &  -4.433 & -0.621 &  -2.426 &  -0.697 &  2.128 &  2.443 &  1.785 & -13.902 & -3.971 &  -8.700 &   0.574 &  0.679 &  0.756 &  1.666 \\
  K &  Br &        RS & -0.166 &  -4.433 & -0.621 &  -2.426 &  -0.697 &  2.128 &  2.443 &  1.785 & -12.650 & -3.739 &  -8.001 &   0.708 &  0.749 &  0.882 &  1.869 \\
  K &   I &        RS & -0.168 &  -4.433 & -0.621 &  -2.426 &  -0.697 &  2.128 &  2.443 &  1.785 & -11.257 & -3.513 &  -7.236 &   0.213 &  0.896 &  1.071 &  1.722 \\
 Ca &   O &        RS & -0.266 &  -6.428 &  0.304 &  -3.864 &  -2.133 &  1.757 &  2.324 &  0.679 & -16.433 & -3.006 &  -9.197 &   2.541 &  0.462 &  0.427 &  2.219 \\
 Ca &   S &        RS & -0.369 &  -6.428 &  0.304 &  -3.864 &  -2.133 &  1.757 &  2.324 &  0.679 & -11.795 & -2.845 &  -7.106 &   0.642 &  0.742 &  0.847 &  2.366 \\
 Ca &  Se &        RS & -0.361 &  -6.428 &  0.304 &  -3.864 &  -2.133 &  1.757 &  2.324 &  0.679 & -10.946 & -2.751 &  -6.654 &   1.316 &  0.798 &  0.952 &  2.177 \\
 Ca &  Te &        RS & -0.350 &  -6.428 &  0.304 &  -3.864 &  -2.133 &  1.757 &  2.324 &  0.679 &  -9.867 & -2.666 &  -6.109 &   0.099 &  0.945 &  1.141 &  1.827 \\
 Cu &   F &        RS & -0.019 &  -8.389 & -1.638 &  -4.856 &  -0.641 &  1.197 &  1.680 &  2.576 & -19.404 & -4.273 & -11.294 &   1.251 &  0.406 &  0.371 &  1.428 \\
 Cu &  Cl &        ZB &  0.156 &  -8.389 & -1.638 &  -4.856 &  -0.641 &  1.197 &  1.680 &  2.576 & -13.902 & -3.971 &  -8.700 &   0.574 &  0.679 &  0.756 &  1.666 \\
 Cu &  Br &        ZB &  0.152 &  -8.389 & -1.638 &  -4.856 &  -0.641 &  1.197 &  1.680 &  2.576 & -12.650 & -3.739 &  -8.001 &   0.708 &  0.749 &  0.882 &  1.869 \\
 Cu &   I &        ZB &  0.203 &  -8.389 & -1.638 &  -4.856 &  -0.641 &  1.197 &  1.680 &  2.576 & -11.257 & -3.513 &  -7.236 &   0.213 &  0.896 &  1.071 &  1.722 \\
 Zn &   O &        ZB &  0.102 & -10.136 &  1.081 &  -6.217 &  -1.194 &  1.099 &  1.547 &  2.254 & -16.433 & -3.006 &  -9.197 &   2.541 &  0.462 &  0.427 &  2.219 \\
 Zn &   S &        ZB &  0.275 & -10.136 &  1.081 &  -6.217 &  -1.194 &  1.099 &  1.547 &  2.254 & -11.795 & -2.845 &  -7.106 &   0.642 &  0.742 &  0.847 &  2.366 \\
 Zn &  Se &        ZB &  0.259 & -10.136 &  1.081 &  -6.217 &  -1.194 &  1.099 &  1.547 &  2.254 & -10.946 & -2.751 &  -6.654 &   1.316 &  0.798 &  0.952 &  2.177 \\
 Zn &  Te &        ZB &  0.241 & -10.136 &  1.081 &  -6.217 &  -1.194 &  1.099 &  1.547 &  2.254 &  -9.867 & -2.666 &  -6.109 &   0.099 &  0.945 &  1.141 &  1.827 \\

 \bottomrule
    \end{tabular}
    }
\end{table}

\begin{table}[h]
    \centering
    \captionsetup{justification=justified,singlelinecheck=false}
    \resizebox{\columnwidth}{!}{%
    \begin{tabular}{lllrrrrrrrrrrrrrrr}

\toprule A &B & \multicolumn{1}{p{2cm}}{\centering DFT \\ Classification} &     $\Delta E$&
  IP(A) &   EA(A) &  HOMO(A) &  LUMO(A) &   rs(A) &   rp(A) &   rd(A) &    IP(B) &   EB(B) &  HOMO(B) &  LUMO(B) &   rs(B) &   rp(B) &   rd(B) \\
\midrule

 Ga &   N &        ZB &  0.433 &  -5.818 & -0.108 &  -2.732 &   0.130 &  0.994 &  1.330 &  2.163 & -13.585 & -1.867 &  -7.239 &   3.057 &  0.539 &  0.511 &  1.540 \\
 Ga &   P &        ZB &  0.341 &  -5.818 & -0.108 &  -2.732 &   0.130 &  0.994 &  1.330 &  2.163 &  -9.751 & -1.920 &  -5.596 &   0.183 &  0.826 &  0.966 &  1.771 \\
 Ga &  As &        ZB &  0.271 &  -5.818 & -0.108 &  -2.732 &   0.130 &  0.994 &  1.330 &  2.163 &  -9.262 & -1.839 &  -5.341 &   0.064 &  0.847 &  1.043 &  2.023 \\
 Ga &  Sb &        ZB &  0.158 &  -5.818 & -0.108 &  -2.732 &   0.130 &  0.994 &  1.330 &  2.163 &  -8.468 & -1.847 &  -4.991 &   0.105 &  1.001 &  1.232 &  2.065 \\
 Ge &  Ge &        ZB &  0.202 &  -7.567 & -0.949 &  -4.046 &   2.175 &  0.917 &  1.162 &  2.373 &  -7.567 & -0.949 &  -4.046 &   2.175 &  0.917 &  1.162 &  2.373 \\

 Rb &   F &        RS & -0.136 &  -4.289 & -0.590 &  -2.360 &  -0.705 &  2.240 &  3.199 &  1.960 & -19.404 & -4.273 & -11.294 &   1.251 &  0.406 &  0.371 &  1.428 \\
 Rb &  Cl &        RS & -0.161 &  -4.289 & -0.590 &  -2.360 &  -0.705 &  2.240 &  3.199 &  1.960 & -13.902 & -3.971 &  -8.700 &   0.574 &  0.679 &  0.756 &  1.666 \\
 Rb &  Br &        RS & -0.164 &  -4.289 & -0.590 &  -2.360 &  -0.705 &  2.240 &  3.199 &  1.960 & -12.650 & -3.739 &  -8.001 &   0.708 &  0.749 &  0.882 &  1.869 \\
 Rb &   I &        RS & -0.169 &  -4.289 & -0.590 &  -2.360 &  -0.705 &  2.240 &  3.199 &  1.960 & -11.257 & -3.513 &  -7.236 &   0.213 &  0.896 &  1.071 &  1.722 \\
Sr &   O &        RS & -0.221 &  -6.032 &  0.343 &  -3.641 &  -1.379 &  1.911 &  2.548 &  1.204 & -16.433 & -3.006 &  -9.197 &   2.541 &  0.462 &  0.427 &  2.219 \\
 Sr &   S &        RS & -0.369 &  -6.032 &  0.343 &  -3.641 &  -1.379 &  1.911 &  2.548 &  1.204 & -11.795 & -2.845 &  -7.106 &   0.642 &  0.742 &  0.847 &  2.366 \\
 Sr &  Se &        RS & -0.375 &  -6.032 &  0.343 &  -3.641 &  -1.379 &  1.911 &  2.548 &  1.204 & -10.946 & -2.751 &  -6.654 &   1.316 &  0.798 &  0.952 &  2.177 \\
 Sr &  Te &        RS & -0.381 &  -6.032 &  0.343 &  -3.641 &  -1.379 &  1.911 &  2.548 &  1.204 &  -9.867 & -2.666 &  -6.109 &   0.099 &  0.945 &  1.141 &  1.827 \\
 Ag &   F &        RS & -0.156 &  -8.058 & -1.667 &  -4.710 &  -0.479 &  1.316 &  1.883 &  2.968 & -19.404 & -4.273 & -11.294 &   1.251 &  0.406 &  0.371 &  1.428 \\
 Ag &  Cl &        RS & -0.044 &  -8.058 & -1.667 &  -4.710 &  -0.479 &  1.316 &  1.883 &  2.968 & -13.902 & -3.971 &  -8.700 &   0.574 &  0.679 &  0.756 &  1.666 \\
 Ag &  Br &        RS & -0.030 &  -8.058 & -1.667 &  -4.710 &  -0.479 &  1.316 &  1.883 &  2.968 & -12.650 & -3.739 &  -8.001 &   0.708 &  0.749 &  0.882 &  1.869 \\
 Ag &   I &        ZB &  0.037 &  -8.058 & -1.667 &  -4.710 &  -0.479 &  1.316 &  1.883 &  2.968 & -11.257 & -3.513 &  -7.236 &   0.213 &  0.896 &  1.071 &  1.722 \\
 Cd &   O &        RS & -0.087 &  -9.581 &  0.839 &  -5.952 &  -1.309 &  1.232 &  1.736 &  2.604 & -16.433 & -3.006 &  -9.197 &   2.541 &  0.462 &  0.427 &  2.219 \\
 Cd &   S &        ZB &  0.070 &  -9.581 &  0.839 &  -5.952 &  -1.309 &  1.232 &  1.736 &  2.604 & -11.795 & -2.845 &  -7.106 &   0.642 &  0.742 &  0.847 &  2.366 \\
 Cd &  Se &        ZB &  0.083 &  -9.581 &  0.839 &  -5.952 &  -1.309 &  1.232 &  1.736 &  2.604 & -10.946 & -2.751 &  -6.654 &   1.316 &  0.798 &  0.952 &  2.177 \\
 Cd &  Te &        ZB &  0.113 &  -9.581 &  0.839 &  -5.952 &  -1.309 &  1.232 &  1.736 &  2.604 &  -9.867 & -2.666 &  -6.109 &   0.099 &  0.945 &  1.141 &  1.827 \\
 In &   N &        ZB &  0.150 &  -5.537 & -0.256 &  -2.697 &   0.368 &  1.134 &  1.498 &  3.108 & -13.585 & -1.867 &  -7.239 &   3.057 &  0.539 &  0.511 &  1.540 \\
 In &   P &        ZB &  0.170 &  -5.537 & -0.256 &  -2.697 &   0.368 &  1.134 &  1.498 &  3.108 &  -9.751 & -1.920 &  -5.596 &   0.183 &  0.826 &  0.966 &  1.771 \\
 In &  As &        ZB &  0.122 &  -5.537 & -0.256 &  -2.697 &   0.368 &  1.134 &  1.498 &  3.108 &  -9.262 & -1.839 &  -5.341 &   0.064 &  0.847 &  1.043 &  2.023 \\
 In &  Sb &        ZB &  0.080 &  -5.537 & -0.256 &  -2.697 &   0.368 &  1.134 &  1.498 &  3.108 &  -8.468 & -1.847 &  -4.991 &   0.105 &  1.001 &  1.232 &  2.065 \\
 Sn &  Sn &        ZB &  0.016 &  -7.043 & -1.039 &  -3.866 &   0.008 &  1.057 &  1.344 &  2.030 &  -7.043 & -1.039 &  -3.866 &   0.008 &  1.057 &  1.344 &  2.030 \\
  B &  Sb &        ZB &  0.581 &  -8.190 & -0.107 &  -3.715 &   2.248 &  0.805 &  0.826 &  1.946 &  -8.468 & -1.847 &  -4.991 &   0.105 &  1.001 &  1.232 &  2.065 \\
 Cs &   F &        RS & -0.112 &  -4.006 & -0.570 &  -2.220 &  -0.548 &  2.464 &  3.164 &  1.974 & -19.404 & -4.273 & -11.294 &   1.251 &  0.406 &  0.371 &  1.428 \\
 Cs &  Cl &        RS & -0.152 &  -4.006 & -0.570 &  -2.220 &  -0.548 &  2.464 &  3.164 &  1.974 & -13.902 & -3.971 &  -8.700 &   0.574 &  0.679 &  0.756 &  1.666 \\
 Cs &  Br &        RS & -0.158 &  -4.006 & -0.570 &  -2.220 &  -0.548 &  2.464 &  3.164 &  1.974 & -12.650 & -3.739 &  -8.001 &   0.708 &  0.749 &  0.882 &  1.869 \\
 Cs &   I &        RS & -0.165 &  -4.006 & -0.570 &  -2.220 &  -0.548 &  2.464 &  3.164 &  1.974 & -11.257 & -3.513 &  -7.236 &   0.213 &  0.896 &  1.071 &  1.722 \\
 Ba &   O &        RS & -0.095 &  -5.516 &  0.278 &  -3.346 &  -2.129 &  2.149 &  2.632 &  1.351 & -16.433 & -3.006 &  -9.197 &   2.541 &  0.462 &  0.427 &  2.219 \\
 Ba &   S &        RS & -0.326 &  -5.516 &  0.278 &  -3.346 &  -2.129 &  2.149 &  2.632 &  1.351 & -11.795 & -2.845 &  -7.106 &   0.642 &  0.742 &  0.847 &  2.366 \\
 Ba &  Se &        RS & -0.350 &  -5.516 &  0.278 &  -3.346 &  -2.129 &  2.149 &  2.632 &  1.351 & -10.946 & -2.751 &  -6.654 &   1.316 &  0.798 &  0.952 &  2.177 \\
 Ba &  Te &        RS & -0.381 &  -5.516 &  0.278 &  -3.346 &  -2.129 &  2.149 &  2.632 &  1.351 &  -9.867 & -2.666 &  -6.109 &   0.099 &  0.945 &  1.141 &  1.827 \\
 Ge &   C &        ZB &  0.808 &  -7.567 & -0.949 &  -4.046 &   2.175 &  0.917 &  1.162 &  2.373 & -10.852 & -0.872 &  -5.416 &   1.992 &  0.644 &  0.630 &  1.631 \\
 Sn &   C &        ZB &  0.450 &  -7.043 & -1.039 &  -3.866 &   0.008 &  1.057 &  1.344 &  2.030 & -10.852 & -0.872 &  -5.416 &   1.992 &  0.644 &  0.630 &  1.631 \\
 Ge &  Si &        ZB &  0.264 &  -7.567 & -0.949 &  -4.046 &   2.175 &  0.917 &  1.162 &  2.373 &  -7.758 & -0.993 &  -4.163 &   0.440 &  0.938 &  1.134 &  1.890 \\
 Sn &  Si &        ZB &  0.136 &  -7.043 & -1.039 &  -3.866 &   0.008 &  1.057 &  1.344 &  2.030 &  -7.758 & -0.993 &  -4.163 &   0.440 &  0.938 &  1.134 &  1.890 \\
 Sn &  Ge &        ZB &  0.087 &  -7.043 & -1.039 &  -3.866 &   0.008 &  1.057 &  1.344 &  2.030 &  -7.567 & -0.949 &  -4.046 &   2.175 &  0.917 &  1.162 &  2.373 \\

 \bottomrule
    \end{tabular}
    }
    \caption{Values related to 82 AB binaries: total energy difference between Rock-Salt and Zinc-Blende ($\Delta E =E^{RS}-E^{ZB}$) (calculated using DFT) and seven atomic properties of  corresponding A and B atom\cite{ghiringhelli_big_2015}. IP, EA, HOMO, LUMO stand for Ionization Potential, Electron Affinity, Highest Occupied Molecular Orbital and Lowest Unoccupied Molecular Orbital, respectively. $rs,rp,rd$ denote distances where the radial probability density reaches the maximum for $s,p,d$ electronic shells, respectively.}
    \label{tab:full_table}
\end{table}

\section{Alloy supercell}
\begin{figure}[t]
 \renewcommand\thefigure{S.4}
    \centering
    \captionsetup{justification=justified,singlelinecheck=false}
    \includegraphics[scale=0.4]{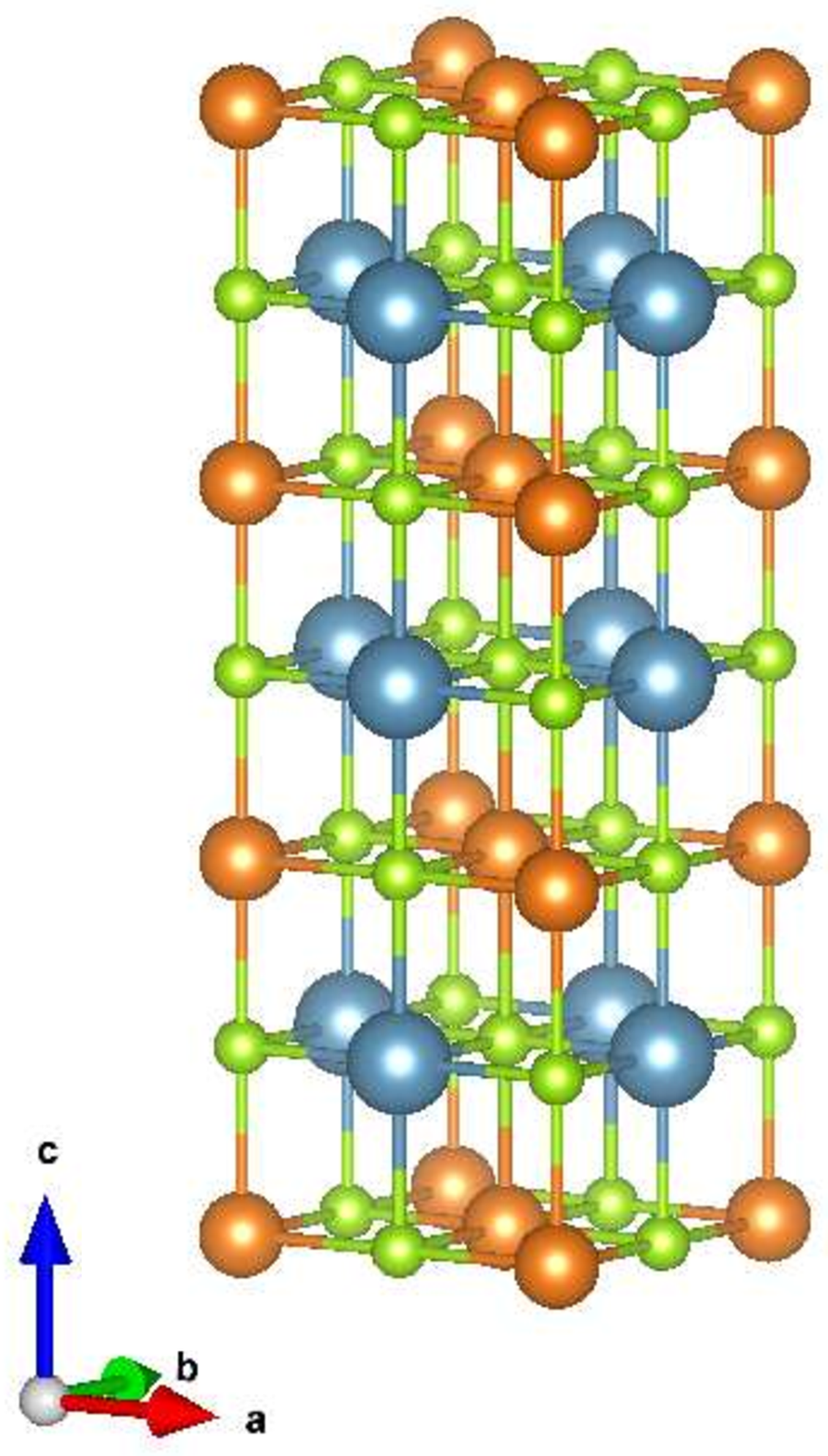}
    \caption{Mg$_{0.5}$Ca$_{0.5}$Se rocksalt supercell: Mg is reported in orange, Ca in blue and Se in green. The supercell is obtained alternating layers of Mg and Ca in the cation sub-lattice along the c primitive vector.}
    \label{fig:supercell}
\end{figure}

\clearpage
\bibliography{RS_ZB_ref.bib}